\documentstyle[fleqn,amsfonts,amssymb]{article}

\def\boxit#1{\vbox{\hrule\hbox{\vrule\kern3pt\vbox{\kern3pt#1\kern3pt%
                                                  }\kern3pt\vrule}\hrule}}
\makeatletter

\newtheorem{theorem}{Theorem}[section]
\newtheorem{lemma}[theorem]{Lemma}
\newtheorem{corollary}[theorem]{Corollary}
\newtheorem{proposition}[theorem]{Proposition}
\newtheorem{algorithm}{Algorithm}[section]
\newtheorem{definition}{Definition}[section]
\newtheorem{definitions}[definition]{Definitions}

\newtheorem{example}{Example}[section]
\newtheorem{exercise}[example]{Exercise}
\newtheorem{actualproblems}[example]{Problems}
\newtheorem{actualexamples}[example]{Examples}
\def\thmstyle{\rm}
\def\indentedtrivlist{\parsep\parskip
  \@trivlist \labelwidth\z@
  \leftmargin\leftmargini  \leftskip\leftmargini
  \expandafter\ifx\csname mathindent\endcsname\relax\else
	\multiply\mathindent by 2
  \fi
  \itemindent\z@ \def\makelabel##1{##1}}
\def\@begintheorem#1#2{\indentedtrivlist \thmstyle
	\item[\hskip-\leftmargini\hskip\labelsep{\bf #1\ #2}]\strut\\\relax}
\def\@opargbegintheorem#1#2#3{\indentedtrivlist \thmstyle
	\item[\hskip-\leftmargini\hskip\labelsep{\bf #1\ #2\ (#3)}]%
	\index{#3}\strut\\\relax}

\def\atendofproof{{\ifvmode\indent\fi
	\unskip\nobreak\hfil\penalty50\vadjust{\penalty500}%
	\hskip2em\mbox{}\nobreak\hfil$\Box$
	\parfillskip=0pt\finalhyphendemerits=0\par}}

\def\listnopb{\let\@beginparpenalty=\@M}

\long\def\titledpar#1{\par\noindent{\bf #1}}
\def\proofsec{\xdef\@ctrname{c@\@currenvir}%
	\expandafter\ifx\csname\@ctrname\endcsname\relax\else
		\par\medskip\rm\fi 
	\@ifnextchar[{\@proofsecarg}{\@proofsecnoarg}}
\def\@proofsecarg[#1]{\titledpar{Proof #1 }}
\def\@proofsecnoarg{\titledpar{Proof: }}

\newenvironment{proof}{\ifvmode \else\par\fi\nobreak\smallskip\nobreak
	\proofsec}{\atendofproof\medbreak}

\newcounter{problem}
\newcounter{examplectr}
\@addtoreset{problem}{example}
\@addtoreset{examplectr}{example}

\def\labelproblem{{\bf\arabic{problem}.}}
\def\labelexamplectr{{\bf(\roman{examplectr})}}
\newenvironment{problems}{\begin{actualproblems}%
	\strut\par\nobreak\global\@nobreaktrue
	\list{\labelproblem}{\usecounter{problem}%
		\def\makelabel##1{\hss\llap{##1}}}\relax
}{\endlist\end{actualproblems}}
\newenvironment{examples}{\begin{actualexamples}%
	\strut\par\nobreak\global\@nobreaktrue
	\list{\labelexamplectr}{\usecounter{examplectr}%
		\def\makelabel##1{\hss\llap{##1}}}\relax
}{\endlist\end{actualexamples}}

\def\bal{\begin{algorithm}}		\def\eal{\end{algorithm}}
\def\ba{\begin{array}}			\def\ea{\end{array}}
\def\bc{\begin{corollary}}		\def\ec{\end{corollary}}
\def\bde{\begin{description}}		\def\ede{\end{description}}
\def\bds{\begin{definitions}}		\def\eds{\end{definitions}}
\def\bd{\begin{definition}}		\def\ed{\end{definition}}
\def\ben{\begin{enumerate}}		\def\een{\end{enumerate}}
\def\be{\begin{eqnarray*}}		\def\ee{\end{eqnarray*}}
\def\bfig{\begin{figure}}		\def\efig{\end{figure}}
\def\bi{\begin{itemize}}		\def\ei{\end{itemize}}
\def\bl{\begin{lemma}}			\def\el{\end{lemma}}
\def\bne{\begin{eqnarray}}		\def\ene{\end{eqnarray}}
\def\bp{\begin{proof}}			\def\ep{\end{proof}}
\def\bse{\begin{equation}}		\def\ese{\end{equation}}
\def\btabb{\begin{tabbing}}		\def\etabb{\end{tabbing}}
\def\btable{\begin{table}}		\def\etable{\end{table}}
\def\btabular{\begin{tabular}}		\def\etabular{\end{tabular}}
\def\bt{\begin{theorem}}		\def\et{\end{theorem}}
\def\bpr{\begin{proposition}}		\def\epr{\end{proposition}}
\def\bver{\begin{verbatim}} 		\def\ever{\end{verbatim}}
\def\bxe{\begin{exercise}}		\def\exe{\end{exercise}}
\def\bxs{\begin{examples}}		\def\exs{\end{examples}}
\def\bx{\begin{example}}		\def\ex{\end{example}}
\def\bprob{\begin{problems}}		\def\eprob{\end{problems}}

	\def\seq#1#2{\ifmmode \{{#1}_{#2}\}\else $\{{#1}_{#2}\}$\fi}
	
	\@namedef{sup*}{\supset}

	\let\oldlor=\lor
	\let\oldland=\land
	\def\lor{\mathrel\oldlor}
	\def\land{\mathrel\oldland}

	\def\set#1for#2\eset{\left\{\:#1\mathbin:#2\:\right\}}

	\def\lcm{\mathop{\rm lcm}\nolimits}

\def\({\left(}				\def\){\right)}
\def\[{\left[}				\def\]{\right]}

\newcount\whichmat \whichmat=0
\newenvironment{pmat}{\whichmat=1 
	\left(\begin{array}}{\end{array}\right)}
\newenvironment{bmat}{\whichmat=2 
	\left[\begin{array}}{\end{array}\right]}
\newenvironment{vmat}{\whichmat=3 
	\left\vert\begin{array}}{\end{array}\right\vert}

\def\m#1#2{
	\ifx(#1 \begin{pmat}{#2}	\else
		\ifx[#1 \begin{bmat}{#2}	\else
			\ifx|#1 \begin{vmat}{#2}	\else
				\begin{#1mat}{#2}
	\fi	\fi	\fi
}

\def\me{
	\ifcase\whichmat
		\errormsg{Error in use of Chris' macros:
			\backslash me when not in a
			matrix environment.}\or
		\end{pmat}\or
		\end{bmat}\or
		\end{vmat}\or
		\end{Vmat}\or
		\end{emat}\or
		\end{cmat}\or
		\end{gmat}\or
		\end{hmat}\or
		\end{qmat}
	\fi
	\whichmat=0
}

\def\thisenumparens#1{%
		\@namedef{the\@enumctr}{(\csname#1\endcsname{\@enumctr})}%
		\@namedef{label\@enumctr}{\csname the\@enumctr\endcsname}%
	}
\def\enumparens#1#2{%
	\@namedef{theenum#1}{(\csname #2\endcsname{enum#1})}%
	\@namedef{labelenum#1}{\csname theenum#1\endcsname}%
	}
\def\thisenumparen#1{%
	\@namedef{the\@enumctr}{\csname #1\endcsname{\@enumctr}}%
	\@namedef{label\@enumctr}{\csname the\@enumctr\endcsname)}%
	}
\def\enumparen#1#2{%
	\@namedef{theenum#1}{\csname #2\endcsname{enum#1}}%
	\@namedef{labelenum#1}{\csname theenum#1\endcsname)}%
	}
\def\thisenumdot#1{%
	\@namedef{the\@enumctr}{\csname #1\endcsname{\@enumctr}}%
	\@namedef{label\@enumctr}{\csname the\@enumctr\endcsname.}%
	}
\def\enumdot#1#2{%
	\@namedef{theenum#1}{\csname #2\endcsname{enum#1}}%
	\@namedef{labelenum#1}{\csname theenum#1\endcsname.}%
	}
\enumdot i{arabic}
\enumparens{ii}{roman}
\enumparen{iii}{alph}

\def\applabel#1{\label{app:#1}}	\def\appref#1{Appendix~\ref{app:#1}}

\def\eqlabel#1{\label{eq:#1}}	\def\eqref#1{(\ref{eq:#1})}
	
\def\figlabel#1{\label{fig:#1}}

\def\seclabel#1{\label{sec:#1}}	\def\secref#1{\S\ref{sec:#1}}
	
\def\thmlabel#1{\label{thm:#1}}	\def\thmref#1{Theorem~\ref{thm:#1}}

\def\bmath#1{\ifmmode\mathpalette\@bmath{#1}\else\@bmath\displaystyle{#1}\fi}
\def\@bmath#1#2{\hbox{\boldmath$%
	\ifx#1\scriptscriptstyle\scriptstyle\else#1\fi
	#2$\unboldmath}}

\expandafter\ifx\csname Bbb\endcsname\relax 
	\let\Bbb\bf

\else 
\fi

\def\BC{\@ifnextchar[\@BCoptarg\@BCnooptarg}
\def\@BCoptarg[#1]{{\Bbb C}^{#1}}
\def\@BCnooptarg{{\Bbb C}}

\def\BN{\@ifnextchar[\@BNoptarg\@BNnooptarg}
\def\@BNoptarg[#1]{{\Bbb N}^{#1}}
\def\@BNnooptarg{{\Bbb N}}

\def\BR{\@ifnextchar[\@BRoptarg\@BRnooptarg}
\def\@BRoptarg[#1]{{\Bbb R}^{#1}}
\def\@BRnooptarg{{\Bbb R}}
\def\BZ{\@ifnextchar[\@BZoptarg\@BZnooptarg}
\def\@BZoptarg[#1]{{\Bbb Z}^{#1}}
\def\@BZnooptarg{{\Bbb Z}}

\def\figps#1#2#3
{
	\begin{figure}
		\centerline{\psfig{figure=#1.ps,height=#2pt}}
		\caption{#3}
		\figlabel{#1}
	\end{figure}
}

\catcode`@=11
\newcount\@@cnt
\def\@@genoneto#1#2{\@@cnt=#1 \advance\@@cnt-1
	\loop \ifnum\@@cnt<#2
		\advance\@@cnt1
		\edef\@@junk{\@@theprefix\number\@@cnt}%
		\input\@@junk
	\repeat
}

\let\@@theprefix\relax
\def\includeupto#1#2#3{\def\@@theprefix{#1}\@@genoneto{#2}{#3}}
\catcode`@=12

\long\outer\def\lnotes#1#2#3#4#5#6#7#8#9
{
	\title
	{
		#2 \\
		#1 Lecture Notes \\
		#3
	}
	\begin{document}
	\expandafter\ifx\csname selectfont\endcsname\relax
	\else
		\family{palatino}\selectfont 
	\fi
	\input{psfig}
	\maketitle
	\pagenumbering{roman}

	This document contains all material for #1, as provided in
	Semester #4 by #3 at the University of Queensland. The
	handbook description of the course is ``#5''.

	#6
	\pagebreak
	\pagenumbering{arabic}
	\includeupto{sec}{#7}{#8}
	\pagebreak
	\bibliographystyle{plain}
	\bibliography{#9}

\begin{thebibliography}{10}

\bibitem{grover1}
L. Grover,
\newblock In {\it Proc. 28th Annual ACM Symposium on the Theory of Computation},
pp. 212-219, ACM Press, New York, 1996

\bibitem{grover2}
L.K. Grover,
\newblock Quantum mechanics helps in searching for a needle in a haystack,
{\it Phys. Rev. Lett.} 79:325, also e-print quant-ph/9706033

\bibitem{knill}
E. Knill,
\newblock On Shor's Quantum Factor Finding Algorithm: Increasing the 
Probability of Success and Tradeoffs Involving the Fourier Transform Modulus,
Technical Report LAUR-95-3350, Los Alamos National Laboratory, August 1995,
also available at 
{\tt http://www.c3.lanl.gov/$\sim$knill/cv/reprints/knill:qc1995c.ps}

\bibitem{shor}
Peter W. Shor
\newblock Polynomial-Time Algorithms for Prime factorization and Discrete 
Logarithms on a Quantum Computer, {\it SIAM J. Comput.}, 26:1484-1509 
(1997)

\bibitem{volovich}
I.V. Volovich,
\newblock Quantum Computing and Shor's Factoring Algorithm, Lectures at 
the Volterra-cirm International School ``Quantum Computer and Quantum 
Information", Trento, Italy, July 25-31, 2001, arXiv:quant-ph/0109004

\bibitem{hardy}
G.H. Hardy and E.M. Wright,
\newblock {\it An Introduction to the Theory of Numbers}, Chap X, 
Oxford University Press and Clarendon Press, Amen House, London, E. C. 4, 
4th edition, published 1960 (original edition published 1938)

\bibitem{knuth}
D.E. Knuth,
\newblock The Art of Computer Programming, Vol 2: Seminumerical Algorithms,
2nd ed., Addison-Wesley, Reading, MA, 1981

\bibitem{remm}
Reinhold Remmert,
\newblock {\it Theory of Complex Functions}, p. 329, Graduate Texts in 
Mathematics, Springer-Verlag, New York, 1991

\bibitem{moore}
Theral O. Moore and Edwin H. Hadlock,
\newblock {\it Complex Analysis}, p. 213, Series in Pure Mathematics, Vol. 9,
World Scientific, Singapore, 1991

\bibitem{beren}
Carlos A. Berenstein and Roger Gay,
\newblock {\it Complex Variables, An Introduction}, p. 225, Graduate Texts in 
Mathematics, Springer-Verlag, New York, 1991

\bibitem{fuchs}
B.A. Fuchs and B.V Shabat, translated by J. Berry and J.W. Reed,
\newblock {\it Functions of a Complex Variable and Some of Their Applications},
p. 331, Permagon Press, 1964

\end{thebibliography}
	\end{document}
}

\makeatother

\title{A Refinement of Shor's Algorithm}

\author{David McAnally\\Distributed Systems Technology Centre\thanks{The work 
reported in this paper has been funded in part by the Co-operative
Research Centre for Enterprise Distributed Systems Technology (DSTC)
through the Australian Federal Government's CRC Programme (Department
of Industry, Science \& Resources)}\\ 
QUT, Brisbane, Qld 4001, Australia\\
Email: {\tt davidm@dstc.edu.au}} 

\begin{document}

\maketitle

\begin{quote}
A refinement of Shor's Algorithm for determining order is introduced, which 
determines a divisor of the order after any one run of a 
quantum computer with almost absolute certainty.  The information garnered 
from each run is accumulated to determine the 
order, and for any $k$ greater than $1$, there is a guaranteed minimum 
positive probability 
that the order will be determined after at most $k$ runs.  The probability 
of determination of the order after at most $k$ runs
exponentially approaches a value negligibly less than one, so that the 
accumulated information determines the order with almost absolute certainty.  
The 
probability of determining the order after at most two runs is more than 60\%,
and the probability of determining the order after at most four runs is more 
than 90\%.
\end{quote}

\section{Introduction}

In quantum computing, there are a few algorithms which can be performed 
more efficiently than their most efficient known classical counterparts.
One such example is Grover's algorithm which improves the efficiency 
of searching an unsorted list to the order of the theoretical limit
of efficiency, at a cost of $O(\sqrt{N})$, where $N$ is the length of 
the list (see for example, \cite{grover1,grover2}).
Another example is supplied by Shor's algorithms for determining order 
and for determining discrete logarithms, both of which can be performed 
in polynomial time with the aid of both a quantum computer and a classical
computer.  A consequence of the fact that Shor's algorithm determines
order in polynomial time is that composite numbers can be factorized in
polynomial time.  Since Shor's algorithms aid in factorizing
composite numbers and in solving the discrete logarithm problem, both in
polynomial time, then their implementation on a quantum computer would 
challenge the security of many of today's
cryptographic algorithms ({\it e.g.} RSA, ElGamal, DSA, ECC).

Shor's original algorithm had the property that the number of runs on the 
quantum computer needed to determine the order of $x$ modulo $n$ was 
$O(\log \log n)$.  In Knill's modification \cite{knill}, the probability of 
success was improved, but on any single run of the quantum computer, the 
probability that the value output by the computers would be a divisor 
of the order may still be significantly less than $1$.  Knill did, however,
introduce the concept of accumulating information from various runs of 
the quantum computer.

It is the purpose of this paper to refine the algorithm to the point that
after any one run on the quantum computer, the probability that the value 
output by the computers is a divisor of the order is negligibly less than 
$1$.  When this refinement is combined with the accumulation of information, 
as discussed above, the number of required runs on the quantum computer is 
reduced to $O(1)$ (assuming ideal working of the quantum computer, 
including extra demands on the Quantum Fourier Transform).  The 
refinement to the algorithm is introduced in \secref{refine}, 
and it is demonstrated in \secref{final} that the probability of 
finding the required order with not more than $k$ runs on the quantum
computer is greater than $1/\zeta(k) - O(n^{-\epsilon})$ in the asymptotic
limit as $n \to \infty$, where $\zeta$ is the Riemann zeta function,
$\epsilon$ is a positive number, and the statement $f > g - O(h)$ means that
there exists a function $F$ such that $f > g - F$ in the asymptotic limit,
and $F/h$ is bounded in the same limit.

The refinement is effected by increasing the number of qubits in the 
first register by a factor of about $1.5$, thus increasing the 
requirements of space and time on the quantum computer by a constant 
factor, and increasing the accuracy required in performing the Quantum 
Fourier Transform on the first register.

In \secref{mod}, the modular metric, which measures distances between elements 
of $\BZ/q\BZ$ is introduced for all $q$.  The purpose for introducing the 
modular metric is in order to obtaining a proper and invariant concept of
proximity.

In \secref{shor}, Shor's original algorithm is discussed.

In \secref{refine}, a refinement of Shor's algorithm is introduced in which 
each run of the quantum computer determines a divisor of the required order
with almost absolute certainty, and the number of required runs on the quantum
computer is $O(1)$.

In \secref{firstreg}, an analysis of the probabilities of the measured value 
of the first register falling in some specific subsets of $\{ 0,1,\dots,q-1 \}$
is given.

In \secref{continued}, some facts about continued fractions (which are used 
in  the classical part of the algorithm to determine information about the 
order) are given, with a new result determining sufficient conditions to
guarantee that the classical part of the algorithm will yield a divisor of the 
required order.

In \secref{algo_analysis}, the results of \secref{firstreg} and 
\secref{continued} are united to demonstrate that the refinement guarantees,
with probability negligibly less than $1$ that each run of the quantum 
yields a divisor of the required order, and the Section also specifies 
sufficient information to determine approximate probabilities for each divisor.

In \secref{prob}, an idealized version of the probability distribution is 
investigated in order to determine the probability that the order will be 
known after at most $k$ runs of the quantum computer.

In \secref{final}, the properties of the idealized probability distribution 
are modified to the more concrete distribution associated with the refinement
of Shor's Algorithm.

\section{Modular Metrics} \seclabel{mod}

For $q \in \BZ$, $q > 1$, let $I_q = \{ 0,1,\dots,q-1 \}$.

\bt \thmlabel{modmet} For $q \in \BZ$, $q > 1$, define 
$\rho_q : I_q \times I_q \to \BZ$ by
\bse \eqlabel{mod}
\rho_q(x,y) = \min(|x-y|,q-|x-y|),
\ese
then $\rho_q$ is a metric on $I_q$, and $\rho_q(x,y) \leq \frac{q}{2}$ for all
$x, y \in I_q$.
\et

This is proven in \appref{modmet}. 

The modular metric $\rho_q$ is equivalent to a metric $s_q$ on $\BZ/q \BZ$
determined by the smallest distance between representatives of the
respective cosets:
\be
s_q(\bar x,\bar y) = \min\{ |x-y| : x \in \bar x, y \in \bar y \},
\ee
for $\bar x, \bar y \in \BZ/q \BZ$.

The modular metric gives a distance function on $\{ 0,1,\dots,q-1 \}$ which 
is invariant under cyclic symmetries, and can be thought of an arc length on 
a circle around which the elements have been evenly spaced.

\section{Shor's Algorithm} \seclabel{shor}

The purpose of the quantum part of Shor's Algorithm
is to determine the order $r$ of $x$ modulo $n$, where
$0 < x < n$, and $x$ and $n$ are relatively prime, in other
words, $r$ is the smallest positive integer such that $x^r \equiv 1
\bmod n$ (note that $0 < r < n$).
In Shor's paper, this was achieved in the following manner.
\ben
\item The state vector of the system is set to an initial state of
\be
| \psi_0 \rangle = \frac{1}{q^{\frac{1}{2}}} \sum_{a=0}^{q-1}
| a \rangle \otimes | 0 \rangle,
\ee
where $q$ is an appropriate power of $2$ (the first register is composed
of $l$ qubits, where $q = 2^l$).  In Shor's paper, $q$ is taken to be
that unique power of $2$ such that $n^2 \leq q < 2 n^2$.  The state vector
$| \psi_0 \rangle$ arises from the state $| \phi_0 \rangle
= | 0 \rangle \otimes | 0 \rangle$ by taking a quantum Fourier transform
on the first register, or alternatively by applying a gate of $H^{\otimes l}$
to the first register (so $H$ is applied individually to each qubit), where
$H$ is the Hadamard gate.
\item The next step is to perform a modular exponentiation, so that
$| \psi_0 \rangle$ is mapped to
\be
| \psi_1 \rangle &=& \frac{1}{q^{\frac{1}{2}}} \sum_{a=0}^{q-1}
| a \rangle \otimes \left| x^a \bmod n \right\rangle\\
&=& \frac{1}{q^{\frac{1}{2}}} \sum_{k=0}^{r-1}
\sum_{b=0}^{\left\lfloor\frac{q-1-k}{r}\right\rfloor}
\left| br+k \right\rangle \otimes \left| x^{br+k} \bmod n \right\rangle\\
&=& \frac{1}{q^{\frac{1}{2}}} \sum_{k=0}^{r-1}
\sum_{b=0}^{\left\lfloor\frac{q-1-k}{r}\right\rfloor}
\left| br+k \right\rangle \otimes \left| x^k \bmod n \right\rangle,
\ee
where for real $y$, $\lfloor y \rfloor$ is the greatest integer less than
or equal to $y$.
The final equality follows from the fact that $r$ is the order of
$x$ modulo $n$.
\item The next step is to take the quantum Fourier transform on the first
register, so that the state becomes
\be
| \psi_2 \rangle &=& \frac{1}{q} \sum_{c=0}^{q-1} \sum_{a=0}^{q-1}
\exp\left(\frac{2\pi iac}{q}\right) \left| c \right\rangle \otimes
\left| x^a \bmod n \right\rangle\\
&=& \frac{1}{q} \sum_{c=0}^{q-1} \sum_{k=0}^{r-1}
\sum_{b=0}^{\left\lfloor\frac{q-1-k}{r}\right\rfloor}
\exp\left(\frac{2\pi i(br+k)c}{q}\right) \left| c \right\rangle \otimes
\left| x^k \bmod n \right\rangle.
\ee
\item The final step is to measure the value $c$ of the first register.
The value of $c$ is then input into a classical computer (which already
has values for $q$ and $n$), and a value for the fraction $d'/r'$ satisfying
the following conditions is found:
\bi
\item $d'/r'$ is in lowest terms ($d'$ and $r'$ have no common factors);
\item $0 \leq d'/r' \leq 1$;
\item $0 < r' < n$;
\item $d'/r'$ is the nearest fraction to $c/q$ which satisfies the other
three conditions.
\ei
This is done with the use of continued fractions.
\een
Shor noted that the probability that $c$ ($c \in \BZ$, $0 \leq c < q$)
is some given value which varies from an integral multiple of $\frac{q}{r}$
by at most $\frac{1}{2}$ (this is equivalent to equation (5.11) of Shor's
paper \cite{shor}), and that the value of the second register is
$x^k \bmod n$ for some given $k$, is greater than $\frac{1}{3r^2}$.
This observation
can be formally expressed as follows: let $X$ be the random variable 
denoting the result of the measurement of the first register, and let $Y$ 
be the random variable denoting the result of a measurement of the second
register, then for any given $d = 0,1,\dots,r-1$ and $k = 0,1,\dots,r-1$, 
\be
P\left(\rho_q\left(X,\frac{dq}{r}\right) \leq \frac{1}{2} \mbox{ and }
Y = x^k \bmod n\right) > \frac{1}{3r^2}.
\ee
It follows that the
probability that $c$ is the value as given above is greater than $\frac{1}{3r}$, and
so the probability that there exists an integer $d$ such that $0 < d < r$,
$d$ is relatively prime to $r$ ($d$ and $r$ have no common factor), and $c$
differs from $\frac{dq}{r}$ by at most $\frac{1}{2}$, is greater than
$\phi(r)/(3r)$, where $\phi$ is Euler's totient function, defined by 
\be
\phi(r) = \#\{ d \in \BZ : 0 < d < r, \ \mbox{$d$ and $r$ are relatively
prime} \}.
\ee
A formula for $\phi$ is given by 
\be
\phi(r) = r \prod_{p \mbox{\small{ prime, }} p | r} \left(1 - \frac{1}{p}
\right).
\ee
The requirement that $d$ and $r$ be relatively prime comes from the fact
that the only information about $d/r$ that can be be derived from $c$ is
its expression in lowest terms (no common factor for numerator and
denominator), so that in order for the denominator to be the order of $x$,
$d$ and $r$ can have no common factors.
Shor used the theorem that $\phi(r)/r > \delta_1/\log \log r$ for some
$\delta_1$ to yield the result that the probability above is greater than
$\delta/\log \log r$, for some $\delta$, so that the number of trials
required on the quantum computer is $O(\log \log n)$.

\section{Refinement of Shor's Algorithm} \seclabel{refine}

The refinement of Shor's Algorithm to be introduced in this paper incorporates
a modification of the value of the parameter $q$, and an accumulation of 
information in a similar manner to that suggested by Knill \cite{knill}.

Take a positive real number $\epsilon$, and let $w = n^\epsilon$.
Under the refinement, the algorithm for determining $r$ is as follows.
All steps except step 2 are performed on a classical computer.
\ben
\item Set $s := 1$ and $q \geq2 w n^3$ ({\it e.g.} set $q$ to
be that unique power of 2 such that $2 w n^3 \leq q < 4 w n^3$);
\item Perform the quantum algorithm on the quantum computer with $q$ as 
specified in Step 1, and measure the value $c$ of the first register;
\item Determine the continued fraction expansion for $\frac{c}{q}$;
\item Determine all denominators of convergents of the continued fraction 
expansion up to the first denominator greater than or equal to $n$;
\item Let $r'$ be the last denominator less than $n$, and set 
$s := \lcm(s,r')$;
\item Calculate $x^s \bmod n$;
\item If $x^s \not\equiv 1 \bmod n$, then go to Step 2;
\item Output $s$.
\een
Note that the algorithm accumulates the information garnered from each
measurement of $c$.  Note also that only the denominators of the convergents 
are calculated.  There is no need to calculate their numerators.

For the size of $n$ that would be typically used in RSA encryption, the 
algorithm above determines the order $r$ with probability negligibly less 
than one (the probability of determining a nontrivial multiple of $r$ instead 
of the correct value is $O(n^{-1})$).  The probability that the correct value 
of $r$ will be found after at most 2 runs of the quantum computer is at least 
60\%, the probability after at most 4 runs is at least 90\%, the probability 
after at most 6 runs is at least 98\%, and the probability after at most 8 
runs is at least 99.5\%.

The rest of the paper is devoted to analysing the above algorithm in order 
to demonstrate the properties claimed for it.

\section{Probabilities of specified values for the first register}
\seclabel{firstreg}

The essential feature of the profile of probabilities of the measured value 
$c$ of the first register is that when $q \gg r$, then the probability 
concentrates in the vicinities of $\frac{dq}{r}$, where $d$ is an integer 
with a probability of about $\frac{1}{r}$ in each vicinity.  Further, if
$\frac{dq}{r}$ is an integer, then the full probability of $\frac{1}{r}$ 
effectively concentrates itself at $c = \frac{dq}{r}$, and if $\frac{dq}{r}$
is not an integer, then the probability of $c$ in the vicinity of 
$\frac{dq}{r}$ is essentially inversely proportional to $(c - \frac{dq}{r})^2$.
It follows that for $q \gg r$, the only dependence that the probability profile
in the vicinity of $\frac{dq}{r}$ has on $q$ is on the fractional part of 
$\frac{dq}{r}$ ({\it i.e.} the full set of profiles is determined completely 
by the fractional part of $\frac{q}{r}$).  This means qualitatively that as 
$q$ increases, the concentrated areas of probability recede from each other,
but the individual profiles do not ``spread".  These observations are made more rigourous in this Section.

All results presented in section without proof will be proven in 
\appref{firstreg}.

Since Shor's Algorithm relies on measuring the value in the first register,
and then entering the result of the measurement into the classical computer,
then it is useful to have information about the probability distribution 
for the values taken by the first register in order to determine the 
probabilities of various outputs of the classical computer.

The parameter $q$ will now be taken to be an arbitrary positive integer,
and a measurement of the first register will be taken when the computer is
in the state
\be
| \psi_2 \rangle = \frac{1}{q} \sum_{c=0}^{q-1} \sum_{k=0}^{r-1}
\sum_{b=0}^{\left\lfloor\frac{q-1-k}{r}\right\rfloor}
\exp\left(\frac{2\pi i(br+k)c}{q}\right) \left| c \right\rangle \otimes
\left| x^k \bmod n \right\rangle.
\ee
Note that $| \psi_2 \rangle$ is the final form of the state vector before
measurement in the quantum algorithm in Shor's algorithm.
The parameter $q$ is generally taken to be a power of $2$ as a result of
the requirement of the usage of qubits in the quantum algorithms for
addition, multiplication and modular exponentiation.  Modification to
qudits (with a higher number of levels) of the algorithms for addition,
multiplication and modular exponentiation will allow for a wider range
of values for $q$.  Also, $q$ is typically taken to be larger than $n$,
although the results below are true for all possible values of $q$.

Let $X$ be the random variable describing the result of the measurement of
the first register in the final step of the algorithm on the quantum
computer, then $X$ must take the value of an integer between $0$ and $q-1$,
inclusive, and for $0 \leq c \leq q-1$, the probability that $X = c$ is
given by $P(X = c) = \langle \chi_c | \chi_c \rangle$, where
\be
\left| \chi_c \right\rangle = \frac{1}{q} \sum_{k=0}^{r-1}
\sum_{b=0}^{\left\lfloor\frac{q-1-k}{r}\right\rfloor}
\exp\left(\frac{2\pi i(br+k)c}{q}\right) \left| x^k \bmod n \right\rangle,
\ee
and so, since $x^k \not\equiv x^{k'} \bmod n$ for $k$ and $k'$ such that
$0 \leq k < r$, $0 \leq k' < r$ and $k \neq k'$, then
\be
P(X = c) &=& \frac{1}{q^2} \sum_{k=0}^{r-1} \left|
\sum_{b=0}^{\left\lfloor\frac{q-1-k}{r}\right\rfloor}
\exp\left(\frac{2\pi i(br+k)c}{q}\right) \right|^2\\
&=& \frac{1}{q^2} \sum_{k=0}^{r-1} \left|
\sum_{b=0}^{\left\lfloor\frac{q-1-k}{r}\right\rfloor}
\exp\left(\frac{2\pi ibcr}{q}\right) \right|^2\\
&=& \frac{1}{q^2} \sum_{k=0}^{r-1} \left|
\sum_{b=0}^{\left\lfloor\frac{q+k}{r}\right\rfloor-1}
\exp\left(\frac{2\pi ibcr}{q}\right) \right|^2,
\ee
where the last equality is obtained by substituting $r-1-k$ for $k$.

If $\frac{cr}{q} \in \BZ$, then
\be
\exp\left(\frac{2\pi icr}{q}\right) = 1,
\ee
so that
\bne
P(X = c) &=& \frac{1}{q^2} \sum_{k=0}^{r-1} \left|
\sum_{b=0}^{\left\lfloor\frac{q+k}{r}\right\rfloor-1} 1 \right|^2 
\eqlabel{cr_int}\\
&=& \frac{1}{q^2} \sum_{k=0}^{r-1} \left\lfloor\frac{q+k}{r}\right\rfloor^2.
\nonumber
\ene

On the other hand, if $\frac{cr}{q} \notin \BZ$, then
\bne
P(X = c) &=& \frac{1}{q^2} \sum_{k=0}^{r-1} \left|
\sum_{b=0}^{\left\lfloor\frac{q+k}{r}\right\rfloor-1}
\exp\left(\frac{2\pi ibcr}{q}\right) \right|^2 \nonumber\\
&=& \frac{1}{q^2} \sum_{k=0}^{r-1} \left|
\frac{\exp\left(\frac{2\pi icr}{q}\left\lfloor\frac{q+k}{r}\right\rfloor
\right)-1}{\exp\left(\frac{2\pi icr}{q}\right)-1} \right|^2 
\eqlabel{cr_non_int}\\
&=& \frac{1}{q^2} \sum_{k=0}^{r-1}
\frac{\sin^2\left(\frac{\pi cr}{q}\left\lfloor\frac{q+k}{r}\right\rfloor
\right)}{\sin^2 \frac{\pi cr}{q}}. \nonumber
\ene
The second equality above follows from the evaluation of the geometric
progression
\be
\sum_{b=0}^{\left\lfloor\frac{q+k}{r}\right\rfloor-1}
\exp\left(\frac{2\pi ibcr}{q}\right)
= \frac{\exp\left(\frac{2\pi icr}{q}\left\lfloor\frac{q+k}{r}\right\rfloor
\right)-1}{\exp\left(\frac{2\pi icr}{q}\right)-1}.
\ee

Much, if not all, of this is already known (e.g. page 17 of \cite{volovich}).

If $\frac{q}{r} \in \BZ$, then
\be
P(X = c) = \left\{ \ba{ll} \frac{1}{r}, & \frac{cr}{q} \in \BZ,\\
0, & \mbox{otherwise}, \ea \right.
\ee
so that $c$ is guaranteed to be a multiple of $\frac{q}{r}$.  Since
$c = \frac{dq}{r}$ for some integer $0 \leq d < r$, then $c/q$ is guaranteed
to be equal to $d/r$ for some $0 \leq d < r$, and all values of $d$ occur
with equal probability $1/r$.

\subsection{The Case That $q/r$ is not an Integer}

The case where $\frac{q}{r} \notin \BZ$ is more difficult.

In the case that $1 \leq q \leq r$, then
\be
P(X = c) = \frac{1}{q},
\ee
for all $c$, so that all possible values of the first register occur with equal
probability, and so no useful information can be obtained, as the behaviour
is independent of $r \geq q$.

Since no useful information can be obtained if $q \leq r$, then from now
it will be assumed that $q > r$.

If $\frac{cr}{q} \in \BZ$, then 
\bse \eqlabel{non_int_int}
\frac{1}{r} - \frac{2}{q} + \frac{r}{q^2} < P(X = c)
< \frac{1}{r} + \frac{2}{q} + \frac{r}{q^2}.
\ese
Note that $P(X = c) = \frac{1}{r} + O(\frac{1}{q})$.
If $q$ is much larger than $r$, then it follows that $P(X = c)$ is very
close to $\frac{1}{r}$.

Suppose $\frac{cr}{q} \notin \BZ$, then 
\bse \eqlabel{upper_bound}
P(X = c) \leq \frac{r}{q^2 \sin^2\frac{\pi cr}{q}}.
\ese
This gives an upper bound for $P(X = c)$, and demonstrates that as the
distance between $c$ and the nearest integral multiple of $\frac{q}{r}$
increases, the maximum possible probability that $X = c$ decreases.  
Specifically, the measured value of the first register is more likely to 
be in the neighbourhood of some multiple of $\frac{q}{r}$ than it is not to 
be in any such neighbourhood.

Suppose that $c = \frac{dq}{r} + \Delta$, where $d \in \BZ$ and $0 <
| \Delta | \leq \frac{q}{2r}$, so that 
\be
P(X = c) &\leq& \frac{r}{q^2 \sin^2\left(\pi d+\frac{\pi \Delta r}
{q}\right)}\\
&=& \frac{r}{q^2 \sin^2\frac{\pi \Delta r}{q}},
\ee
by straightforward substitution for $c$ in \eqref{upper_bound}.

If $\frac{dq}{r} \in \BZ$, so that $\Delta \in \BZ$, then
\bse \eqlabel{non_int_int_delta}
P(X = c) < \frac{r}{q^2}.
\ese
Since $P(X = c) = O(\frac{r}{q^2})$, then for $q$ much larger than $r$, 
$P(X = c)$ is approximately equal to zero.

If $\frac{dq}{r} \notin \BZ$, so that $\Delta \notin \BZ$, then
\bse \eqlabel{delta_upper}
P(X = c) < \frac{1}{\left(1 - \frac{\pi^2 \Delta^2 r^2}{6 q^2}\right)^2}
\left(\frac{\sin^2\frac{\pi dq}{r}}{\pi^2 \Delta^2 r}
+ \frac{2 \left|\sin\frac{\pi dq}{r}\right|}{\pi |\Delta| q}
+ \frac{r}{q^2}\right).
\ese
Further, if $|\Delta| \leq \frac{q}{\pi r} |\sin\frac{\pi dq}{r}|$, then
\bse \eqlabel{delta_lower}
P(X = c) > \frac{\sin^2\frac{\pi dq}{r}}{\pi^2 \Delta^2 r}
- \frac{2 \left|\sin\frac{\pi dq}{r}\right|}{\pi |\Delta| q} + \frac{r}{q^2}.
\ese
It follows that 
\be
P(X = c) = \frac{\sin^2\frac{\pi dq}{r}}{\pi^2 \Delta^2 r} 
+ O\left(\frac{1}{q}\right),
\ee
so that if $q$ is much larger than $r$, then
\be
P(X = c) \sim \frac{\sin^2\frac{\pi dq}{r}}{\pi^2 \Delta^2 r},
\ee
and so $P(X = c)$ is inversely proportional to $(c - \frac{dq}{r})^2$ 
in the asymptotic limit.  Note that the asymptotic profile is dependent only
on the fractional part of $\frac{dq}{r}$, and not on the size of $q$.

\subsection{Probabilities for Certain Subsets}

Since the fraction which interests us, as far as determining the order is 
concerned, is not $\frac{c}{q}$ (where $c$ is the measured value of the 
first register), but $\frac{d}{r}$, then the probability that $\frac{c}{q}$ 
falls in the proximity of $\frac{d}{r}$ is important, and the probability 
that $\frac{c}{q}$ falls within a certain distance of $\frac{d}{r}$ (or,
equivalently, that $c$ falls within a certain distance of $\frac{dq}{r}$),
will be determined for a certain range of distances.

If $\frac{dq}{r} \in \BZ$, then
\bse \eqlabel{int_full}
\frac{1}{r} - \frac{2}{q} + \frac{r}{q^2}
< P\left(\rho_q\left(X,\frac{dq}{r}\right) \leq \frac{q}{2r} \right)
< \frac{1}{r} + \frac{3}{q} + \frac{r}{q^2},
\ese
where $\rho_q$ is the modular metric \eqref{mod}.
Note that $P(\rho_q(X,\frac{dq}{r}) \leq \frac{q}{2r}) = \frac{1}{r}
+ O(\frac{1}{q})$.
If $q$ is much larger than $r$, then it follows that
$P(\rho_q(X,\frac{dq}{r}) \leq \frac{q}{2r})$ is very close to $\frac{1}{r}$.

The value of the parameter $q$ will now be restricted so that $q > 2r$.

From now, for $c \in \{ 0,1,\dots,q-1 \}$, $d_c$ and $\Delta_c$ will be
uniquely determined by the following conditions:
\ben
\item $d_c \in \{ 0,1,\dots,r \}$;
\item $c = \frac{d_c q}{r} + \Delta_c$;
\item $- \frac{q}{2r} < \Delta_c \leq \frac{q}{2r}$.
\een

For $0 < u \leq \frac{q}{2r} - 1$,
\bse \eqlabel{far}
P(|\Delta_X| \geq u + 1) < \frac{2}{\pi^2 u}.
\ese
This determines a hard upper bound independent of $q$ for the probability
that the distance between the measured value of the first register and the
nearest multiple of $\frac{q}{r}$ exceeds any given value greater than $1$
but no greater than $\frac{q}{2r}$, and demonstrates that the measured value
of the first register will tend to be close to a multiple of $\frac{q}{r}$.
Specifically, for any large fixed distance, the probability that the difference 
between the measured value of the first register and the nearest multiple 
of $\frac{q}{r}$ exceeds this distance is small, independent of the size 
of $q$.

Let $u$ now be fixed subject to $0 < u \leq \frac{q}{2r} - 1$.

If $\frac{dq}{r} \in \BZ$, then
\bse \eqlabel{int}
\frac{1}{r} - \frac{2}{q} + \frac{r}{q^2}
< P\left(\rho_q\left(X,\frac{dq}{r}\right) < u + 1\right)
< \frac{1}{r} + \frac{2}{q} + \frac{(2u+3)r}{q^2}.
\ese

If $\frac{dq}{r} \notin \BZ$, then it follows from the bounds already 
determined on $P(X = c)$ for $c$ such that 
\be
\left|c - \frac{dq}{r}\right| < u + 1,
\ee
that if 
\be
u \leq \frac{q}{\pi r} \left| \sin\frac{\pi dq}{r} \right| - 1,
\ee
then
\bne
&& \frac{1}{r} - \frac{2}{\pi^2 r u} - \frac{2}{\pi q} \left(\frac{r^2}{r-1}
+ \ln\frac{r^2(u+1)^2}{r-1}\right) + \frac{r (2u+1)}{q^2} \nonumber\\
&<& P\left(\left| X - \frac{dq}{r} \right| < u+1\right) \eqlabel{non_int}\\
&<& \frac{1}{\left(1 - \frac{\pi^2 (u+1)^2 r^2}{6 q^2}\right)^2}
\left(\frac{1}{r} + \frac{2}{\pi q} \left(\frac{r^2}{r-1}
+ \ln\frac{r^2(u+1)^2}{r-1}\right) + \frac{r (2u+3)}{q^2}\right). \nonumber
\ene
Note that if $u$ is large, and if $q$ is much larger than $r u$,
then $P(|X-\frac{dq}{r}| < u+1)$ is very close to $\frac{1}{r}$.

The probability that the measured value of the first register will be in the
proximity of any specified multiple of $\frac{q}{r}$ has been determined 
to be very close to $\frac{1}{r}$ for any given multiple, and so the 
probability that $\frac{c}{q}$ (where $c$ is the measured value of the 
first register) is close to $\frac{d}{r}$ is approximately $\frac{1}{r}$
for any given value of $d$.

In summary, for $q$ very large, the nett probability of $1$ is equally divided 
amongst the vicinities of $\frac{dq}{r}$ for $d \in \BZ$, with the probability
effectively concentrated within vicinities of fixed maximum width, so that 
as $q$ increases, the vicinities recede from each other while maintaining 
their maximum widths.

\section{Continued Fractions} \seclabel{continued}

The determination of an appropriate rational number $\frac{d'}{r'}$
from the measured value of $c$ is done on a classical computer with the
use of continued fractions (see \cite{shor}, for example).  
In the context of the refinement of Shor's Algorithm, we are interested in 
the width of the vicinity of $\frac{dq}{r}$ which will, with certainty, 
identify $\frac{d}{r}$ as the correct approximation to $\frac{c}{q}$ 
where $c$ is the measured value of the first register.  The width is linearly
dependent on $q$.

The definition of a continued fraction is given here, along with some useful 
properties.

The definition of continued fractions and most of the consequences, as drawn 
below, can be found in \cite{hardy} and \cite{knuth}.

For integers $a_0,a_1,a_2,\dots,a_N$, where $a_0 \geq 0$ and $a_i > 0$
for $i = 1,2,\dots,N$, define the continued fraction $[a_0,a_1,a_2,\dots,
a_N]$ by
\be
[a_0,a_1,a_2,\dots,a_N] = a_0 + \frac{1}{a_1+\frac{1}{a_2+\frac{1}{\dots
+\frac{1}{a_N}}}},
\ee
so that $[a_0,a_1,a_2,\dots,a_N]$ is a rational number.  Alternatively,
a finite continued fraction can be defined by induction on the number of
terms as follows.  For a non-negative integer $a_0$, define $[a_0] = a_0$,
and for integers $a_0,a_1,a_2,\dots,a_N$, as above, define
\be
[a_0,a_1,a_2,\dots,a_N] = a_0 + \frac{1}{[a_1,a_2,\dots,a_N]}.
\ee
For any $0 \leq k \leq N$, $\xi_k = [a_0,a_1,\dots,a_k]$ is called a
convergent of the continued fraction expansion.

If $\xi = \frac{p}{q}$ is rational, then define $a_i$ and $\zeta_i$ by 
induction on $i$ by
\be
\zeta_i &=& \left\{ \ba{ll} \xi, & i = 0,\\ \frac{1}{\zeta_{i-1}-a_{i-1}}, &
\mbox{otherwise, if $\zeta_{i-1} \ne a_{i-1}$}, \ea \right.\\
a_i &=& \lfloor \zeta_i \rfloor,
\ee
terminating when $\zeta_i = a_i$ ({\it i.e.} when $\zeta_i$ is an integer).
This gives a continued fraction expansion $\xi = [a_0,a_1,a_2,\dots,a_N]$, 
where $a_N > 1$.
Alternatively, $\xi = [a_0,a_1,a_2,\dots,a_N-1,1]$, yielding two distinct
continued fraction expansions for $\xi$.
It is known that for any rational number $\xi$, these two continued fraction
expansions are the only possible expansions (for irrational numbers, there is 
exactly one continued fraction expansion, which is infinite).

Define integers $p_k$ and $q_k$ for $k \geq -1$ by induction on $k$ as
follows.
Let 
\bne
p_k &=& \left\{ \ba{ll} 1, & k = -1,\\a_0, & k = 0,\\
p_{k-2} + a_k p_{k-1}, & k > 0, \ea \right. \eqlabel{numer}\\
q_k &=& \left\{ \ba{ll} 0, & k = -1,\\1, & k = 0,\\
q_{k-2} + a_k q_{k-1}, & k > 0, \ea\right. \eqlabel{denom}
\ene
then by standard results from the theory of continued fractions,
\bi
\item $\xi_k = p_k/q_k$ for $k = 0,1,2,\dots$;
\item $p_{k-1} q_k - p_k q_{k-1} = (-1)^k$ for $k = 0,1,2,\dots$;
\item $\xi_{k+1} - \xi_k = \frac{(-1)^k}{q_k q_{k+1}}$;
\item $\gcd(p_k,q_k) = 1$.
\ei
The first two statements are very easily proved by induction, and the
third and fourth statements are trivial consequences of the first two.

It is also well-known that if $\xi$ is a positive real number, $p$ and $q$ 
are positive integers, and $|\xi - \frac{p}{q}| \leq \frac{1}{2q^2}$, then 
$\frac{p}{q}$ is a convergent of the continued fraction expansion for $\xi$ 
(see for example, \cite{hardy,knuth}).

It is proven in \cite{hardy} that
\bt \thmlabel{hardy} For $k > 1$, let $\frac{p_k}{q_k}$ be the corresponding 
convergent of the continued fraction expansion for $\xi$, so that $p_k$ and 
$q_k$ are defined by \eqref{numer} and \eqref{denom}, then for
$0 < q \leq q_k$ and $p \in \BZ$ such that $\frac{p}{q} \ne \frac{p_k}{q_k}$, 
\be
| p - q \xi | \geq | p_k - q_k \xi |,
\ee
and
\be
\left| \xi - \frac{p}{q} \right| > \left| \xi - \frac{p_k}{q_k} \right|.
\ee
\et

We now come to the principal result that will be of use in analysing the 
refinement of Shor's Algorithm, since it gives a sufficient condition on 
$c$ (the result of measuring the first register) that will guarantee that 
the nearest fraction to $\frac{c}{q}$ with denominator less than $n$ is 
$\frac{d}{r}$ for some integer $d$, and that $\frac{d}{r}$ is a convergent 
of the continued fraction expansion for $\frac{c}{q}$.
\bt \thmlabel{square} Suppose $r,n \in \BZ$ and $0 < r < n$.
For $v > 1$, let $q$ be an integer greater than or equal to
$2 v n^2$.  Suppose $d \in \{ 0,1,\dots,r \}$ and $|c-\frac{dq}{r}| \leq v$.
Let $d'/r'$ be the fraction satisfying the following conditions:
\bi
\item $d'/r'$ is in lowest terms ($d'$ and $r'$ have no common factors);
\item $0 \leq d'/r' \leq 1$;
\item $0 < r' < n$;
\item $d'/r'$ is the nearest fraction to $c/q$ which satisfies the other
three conditions.
\ei
Then $d'/r' = d/r$, and $d/r$ is a convergent of the continued fraction
expansion for $c/q$.
Define $p_k$ and $q_k$ for $k = 0,1,\dots$, by \eqref{numer} and \eqref{denom},
respectively.  Let $N = \max \{ k : q_k < n \}$, then
$\xi_N = p_N/q_N = d/r$, so that $d/r$ is the last convergent of the
expansion which has denominator less than $n$.
\et
\bp Since $|c-\frac{dq}{r}| \leq v$, then
\be
\left|\frac{c}{q} - \frac{d}{r}\right| \leq \frac{v}{q} \leq \frac{1}{2n^2}
< \frac{1}{2r^2},
\ee
so that $d/r$ is a convergent of the continued fraction expansion for $c/q$.

Suppose $f,s \in \BZ$, $0 < s < n$, and $0 \leq f \leq s$.  If $f/s \ne d/r$,
then
\be
\left|\frac{f}{s} - \frac{d}{r}\right| = \left|\frac{fr-ds}{rs}\right|
\geq \frac{1}{rs} > \frac{1}{n^2},
\ee
so that
\be
\left|\frac{c}{q} - \frac{f}{s}\right| \geq
\left|\frac{d}{r} - \frac{f}{s}\right|
- \left|\frac{c}{q} - \frac{d}{r}\right|
> \frac{1}{n^2} - \frac{1}{2n^2} = \frac{1}{2n^2}
\geq \left|\frac{c}{q} - \frac{d}{r}\right|,
\ee
so that $d/r$ is the nearest fraction to $c/q$ which satisfies the requisite 
three conditions.

Since $d/r = d'/r'$ is a convergent of the continued fraction expansion for 
$c/q$, then there exists $N$ such that $p_N = d'$ and $q_N = r' < n$.
If $q_{N+1} < n$, then, as a consequence of \thmref{hardy}, 
\be
\left|\frac{c}{q} - \frac{p_{N+1}}{q_{N+1}}\right| 
< \left|\frac{c}{q} - \frac{p_N}{q_N}\right|
= \left|\frac{c}{q} - \frac{d'}{r'}\right|,
\ee
contradicting the fact that $d/r$ is the nearest fraction to $c/q$ with 
denominator less than $n$.  It follows that $q_{N+1} \geq n$, and so $d/r$ 
is the last convergent of the expansion which has denominator less than $n$.
\ep

\section{Some Analysis of the Refinement of Shor's algorithm}
\seclabel{algo_analysis}

Recall the refinement of the Shor's Algorithm as given earlier.
All steps except step 2 are performed on a classical computer.
\ben
\item Set $s := 1$ and $q \geq 2 w n^3$ ({\it e.g.} set $q$ to
be that unique power of 2 such that $2 w n^3 \leq q < 4 w n^3$);
\item Perform the quantum algorithm on the quantum computer with $q$ as 
specified in Step 1, and measure the value $c$ of the first register;
\item Determine the continued fraction expansion for $\frac{c}{q}$;
\item Determine all denominators of convergents of the continued fraction 
expansion up to the first denominator greater than or equal to $n$;
\item Let $r'$ be the last denominator less than $n$, and set 
$s := \lcm(s,r')$;
\item Calculate $x^s \bmod n$;
\item If $x^s \not\equiv 1 \bmod n$, then go to Step 2;
\item Output $s$.
\een

The significant results of the last two sections can be summarised as follows:
\bi
\item For $q$ very large, the nett probability of $1$ is essentially equally 
divided amongst vicinities of $\frac{dq}{r}$ of fixed finite maximum width for 
$d \in \BZ$;
\item The width of the vicinity of $\frac{dq}{r}$ which will, with certainty, 
identify $\frac{d}{r}$ as the correct approximation to $\frac{c}{q}$,
is linearly dependent on $q$.
\ei
This means that if a large enough value for $q$ is taken, then the vicinity 
which will, with certainty, identify $\frac{d}{r}$ as the correct approximation
to $\frac{c}{q}$, will encompass the entire vicinity of $\frac{dq}{r}$ in which 
the probability is effectively concentrated.  This is the {\it raison d'\^etre}
for choosing $q$ with the value as given in the refinement.

In the refinement of Shor's Algorithm, then
\be
P(|\Delta_X| \geq w n) < \frac{2}{\pi^2 (w n - 1)},
\ee
as a consequence of \eqref{far}, so that
\bse
P(|\Delta_X| < w n) > 1 - \frac{2}{\pi^2 (w n - 1)}, \eqlabel{less}
\ese
and so if $n$ is large, then $P(|\Delta_X| \geq w n)$ is very small, and
$P(|\Delta_X| < w n)$ is very close to 1.  If $\frac{dq}{r} \in \BZ$, then
by \eqref{int},
\bse \eqlabel{ref1}
\frac{1}{r} - \frac{1}{w n^3}
< P\left(\rho_q\left(X,\frac{dq}{r}\right) < w n\right)
< \frac{1}{r} + \frac{1}{w n^3} + \frac{1}{2 w n^4}.
\ese
This result is proven in \appref{algo_analysis}.

If $\frac{dq}{r} \notin \BZ$, then, by \eqref{non_int},
\bne
&& \frac{1}{r} - \frac{2}{\pi^2 (wn - 1)} - \frac{1}{\pi w n^3}
\left(n + 1 + \ln\frac{w^2 n^4}{n-1}\right) \nonumber\\
&<& P\left(\left| X - \frac{dq}{r} \right| < w n\right) \eqlabel{ref2}\\
&<& \frac{1}{\left(1 - \frac{\pi^2}{24 n^2}\right)^2}
\left(\frac{1}{r} + \frac{1}{\pi w n^3} \left(n + 1
+ \ln\frac{w^2 n^4}{n-1}\right) + \frac{1}{2 w n^4}\right). \nonumber
\ene
This result is also proven in \appref{algo_analysis}.

It follows that if $n$ is large, as in the case for any practical RSA
encryption algorithm, then $P(\rho_q(\Delta_X,\frac{dq}{r}) < w n)$ is
close to $1/r$ for all $d$.

For the refinement, the probability that $|\Delta_c| < wn$,
where $c$ is the measured value of the first register, is greater than 
$1 - \frac{2}{\pi^2 (wn - 1)}$ by \eqref{less}.  By \thmref{square}, 
if $|\Delta_c| < wn$, then the last convergent of the continued fraction 
expansion for $c/q$ with denominator less than $n$ is necessarily of the 
form $d/r$ for some $d \in \BZ$ such that $0\leq d \leq r$, so that, in 
the refinement, $r'$ necessarily divides $r$, as this is the convergent 
which is determined by the  refinement (or rather, its denominator is 
determined by the refinement).  
It follows that after each run on the quantum computer, the probability 
that $r'$ divides $r$ is greater than $1 - \frac{2}{\pi^2 (wn - 1)}$.  
Since the runs on the quantum computer are, in effect, independent random 
samples with replacement, then the probability that $s$ still divides $r$ 
after $k$ runs on the quantum computer is greater than 
$(1 - \frac{2}{\pi^2 (wn - 1)})^k$.  Specifically, for the size of $n$ that 
would typically be used in RSA encryption, the probability that $s$ will not 
divide $r$ after $k$ runs on the quantum computer is negligibly small (of the 
same order of magnitude as $\frac{k}{wn}$).  Since the value of $s$ is 
almost guaranteed to be a divisor of $r$ after $k$ runs of the quantum 
computer, and $x^s \equiv 1 \bmod n$ iff $s$ is a multiple of $r$, then 
it is almost guaranteed that when the refinement terminates, $s$ will be equal 
to $r$ ($s$ is certainly a multiple of $r$ on termination, and it is almost 
certain to be a divisor of $r$).

This can be expressed formally as follows.  Let $A_k$ denote the random 
variable describing the result of the measurement of the first register 
after the $k$-th run of the quantum computer, let $B_k$ denote the random
variable describing the corresponding value of $r'$ calculated by the 
classical computer, and let $C_k$ be the random variable defined by 
\be
C_k = \lcm(B_1,\dots,B_k),
\ee
so that $C_k$ describes the value of $s$ after $k$ runs of the quantum 
computer, then, by \eqref{less},
\be
P(|\Delta_{A_k}| < w n) > 1 - \frac{2}{\pi^2 (w n - 1)},
\ee
for all $k$, so that by \thmref{square},
\be
P(B_k | r) > 1 - \frac{2}{\pi^2 (w n - 1)},
\ee
for all $k$, and so
\be
P(C_k | r) > \left(1 - \frac{2}{\pi^2 (w n - 1)}\right)^k,
\ee
for all $k$, by the independence of the random variables $A_k$ (from which
the independence of the random variables $B_k$ follows).

\section{Some Results in Probability} \seclabel{prob}

It was noted in \secref{algo_analysis} that the probability that $\frac{d}{r}$ 
is the fraction with denominator less than $n$ which is closest to 
$\frac{c}{q}$ (where $c$ is the measured value of the first register) is close 
to $\frac{1}{r}$ (and in fact, it approaches $\frac{1}{r}$ in the limit as 
$q \to \infty$).  The properties of the probability distributions of certain 
random variables (which are analogues of important random variables related 
to the refinement) associated with the idealized distribution follow.
The purpose here is to get some idea of the probability that the refinement
of Shor's Algorithm will terminate after at most $k$ runs of the quantum 
computer and output the required order.

Let a natural number $s$ have prime factorization 
\be
s = \prod_{j \in J} p_j^{a_j},
\ee
where $J$ is some index set, $p_j$ are distinct primes, and $a_j \geq 1$ 
for all $j \in J$.
Let $Z_i, \ i = 1,2,3,\dots$, denote independent uniformly distributed 
random variables from the sample space $\{ 0,1,\dots,s-1 \}$, so that for 
all $i$, and for all $d$ in the sample space, $P(Z_i = d) = \frac{1}{s}$.  
Let $R_i$ be the random variable defined by 
\be
R_i = \frac{s}{\gcd(Z_i,s)},
\ee
so that $R_i$ are independent random variables, and $R_i$ is the denominator 
of $\frac{Z_i}{s}$, when expressed in lowest terms.
For $k = 1,2,3,\dots$, define the random variable $S_k$ by 
\be
S_k = \lcm(R_1,R_2,\dots,R_k).
\ee
Note that $s$ is a parameter for the probability distributions of $Z_i$,
$R_i$, and $S_k$.
\bt \thmlabel{zeta} For all values of the parameter $s$, 
\be
P(S_k = s) > \frac{1}{\zeta(k)},
\ee
for all $k \geq 2$, where $\zeta$ is the Riemann zeta function defined 
by
\be
\zeta(z) = \sum_{n=1}^\infty \frac{1}{n^z} = \prod_{p \mbox{\small{ prime}}} 
\frac{1}{1 - \frac{1}{p^z}},
\ee
for $\Re(z) > 1$.
\et
Specifically,
\be
P(S_2 = s) > \frac{6}{\pi^2} > \frac{3}{5}, \quad
P(S_4 = s) > \frac{90}{\pi^4} > \frac{9}{10},
\ee
so that the probability that $S_2$ is equal to $s$ is greater than 60\%, 
and the probability that $S_4$ is equal to $s$ is greater than 90\%.
Similarly,
\be
P(S_6 = s) > \frac{945}{\pi^6} > \frac{49}{50}, \quad
P(S_8 = s) > \frac{9450}{\pi^8} > \frac{199}{200},
\ee
so that the probability that $S_6$ is equal to $s$ is greater than 98\%, 
and the probability that $S_8$ is equal to $s$ is greater than 99.5\%.

The proof of \thmref{zeta} is given in \appref{zeta}

\section{More Analysis of the Refinement of Shor's Algorithm} \seclabel{final}

As before, let $A_k$ denote the random variable describing the 
result of the measurement of the first register after the $k$-th 
run of the quantum computer, let $B_k$ describe the corresponding 
value of $r'$ as determined by the refinement, and let the random variable
$C_k$ be defined by
\be
C_k = \lcm(B_1,\dots,B_k).
\ee
Further, let the random variable $D_k$ be defined by
\be
D_k = \left\lfloor \frac{rA_k}{q} + \frac{1}{2} \right\rfloor,
\ee
so that $D_k$ describes the nearest integer to $\frac{rc}{q}$, where 
$c$ is the measured value of the first register after the $k$-th run 
of the quantum computer.  Note that $A_k, \ k = 1,2,3,\dots$, are 
independent random variables, and that for $c = 0,\dots,q-1$,
\be
P(A_k = c) = \left\{ \ba{ll} 
\frac{1}{q^2} \sum_{m=0}^{r-1} \left\lfloor\frac{q+m}{r}\right\rfloor^2,
& \frac{cr}{q} \in \BZ,\\
\\
\frac{1}{q^2} \sum_{m=0}^{r-1}
\frac{\sin^2\left(\frac{\pi cr}{q}\left\lfloor\frac{q+m}{r}\right\rfloor
\right)}{\sin^2 \frac{\pi cr}{q}}, & \frac{cr}{q} \notin \BZ, \ea \right.
\ee
and as noted previously, for the refinement,
\be
P(|\Delta_{A_k}| < w n) > 1 - \frac{2}{\pi^2 (w n - 1)},
\ee
so that 
\be
P(B_k | r) > 1 - \frac{2}{\pi^2 (w n - 1)}, 
\ee
by \thmref{square}, and so
\be
P(C_k | r) > \left(1 - \frac{2}{\pi^2 (w n - 1)}\right)^k.
\ee

It follows that for any given $d$, by \eqref{ref1} and \eqref{ref2},
\be
\frac{1}{r} - \frac{1}{w n^3}
< P\left(\rho_q\left(A_k,\frac{dq}{r}\right) < w n\right) 
< \frac{1}{r} + \frac{1}{w n^3} + \frac{1}{2 w n^4},
\ee
if $\frac{dq}{r} \in \BZ$, and
\be
&& \frac{1}{r} - \frac{2}{\pi^2 (wn - 1)} - \frac{1}{\pi w n^3}
\left(n + 1 + \ln\frac{w^2 n^4}{n-1}\right)\\
&<& P\left(\left| A_k - \frac{dq}{r} \right| < w n\right)\\
&<& \frac{1}{\left(1 - \frac{\pi^2}{24 n^2}\right)^2}
\left(\frac{1}{r} + \frac{1}{\pi w n^3} \left(n + 1
+ \ln\frac{w^2 n^4}{n-1}\right) + \frac{1}{2 w n^4}\right),
\ee
if $\frac{dq}{r} \notin \BZ$.

This concludes the summary of what is already known.

Note that $D_k = d$ if 
\be
\rho_q\left(A_k,\frac{dq}{r}\right) < wn,
\ee
for $d = 1,\dots,r-1$, and $D_k = 0$ or $D_k = r$ if
\be
\rho_q(A_k,0) < wn.
\ee

Given $\rho_q(A_k,\frac{dq}{r}) < wn$, for some $d = 0,1,\dots,r$, then 
$D_k = d$, and 
\be
B_k = \frac{r}{\gcd(D_k,r)}.
\ee
It follows that for all $d$,
\be
P\left(\rho_q\left(A_k,\frac{dq}{r}\right) < wn \mbox{ and } 
D_k = d\right) = \frac{1}{r} + O\left(\frac{1}{wn}\right),
\ee
and 
\be
P\left( \rho_q\left(A_k,\frac{dq}{r}\right) \geq wn \mbox{ for all } d \right)  
= O\left(\frac{1}{wn}\right).
\ee
This means that the probability distribution for $D_k$ becomes uniform 
in the asymptotic limit, and the results of the last Section become exact 
in the asymptotic limit.  Here, $D_k$ plays the same role as $Z_i$, 
$B_k$ plays the same role as $R_i$, and $C_k$ plays the same role as 
$S_k$.  This means that the asymptotic limit of $P(C_k = r)$ as $n$ becomes
large should be greater than $\frac{1}{\zeta(k)}$ for $k \geq 2$.

By a similar argument to that used in the proof of \thmref{zeta} (in 
\appref{zeta}), for $k \geq 2$ and 
$k$ small, 
\bne
P(C_k = r) &=& \prod_{j \in J} \left(1 - \frac{1}{p_j^k}\right) 
+ O\left(\frac{1}{w}\right) \nonumber\\
&=& \prod_{j \in J} \left(1 - \frac{1}{p_j^k}\right) 
+ O(n^{-\epsilon}), \eqlabel{ck=r}
\ene
since $w = n^\epsilon$.
Finally, since 
\be
\prod_{j \in J} \left(1 - \frac{1}{p_j^k}\right) 
> \prod_{p \mbox{\small{ prime}}} \left(1 - \frac{1}{p^k}\right) 
= \frac{1}{\zeta(k)},
\ee
then
\be
P(C_k = r) > \frac{1}{\zeta(k)} - O(n^{-\epsilon}),
\ee
where the statement $f > g - O(h)$ means that there exists a function 
$F$ such that $f > g - F$ in the asymptotic limit, and $F/h$ is bounded 
in the same limit.

The derivation of \eqref{ck=r} is given in \appref{prob}.

This means that for the size of $n$ that would typically be used in RSA 
encryption, the probability that the correct value for $r$ will be found 
after at most $2$ runs of the quantum computer is at least 60\%, and 
the probability that the correct value for $r$ will be found 
after at most $4$ runs of the quantum computer is at least 90\%, {\it etc}.

\section{Conclusion}

There are various advantages and disadvantages to the refinement of 
Shor's algorithm as detailed in this paper.  The advantages include 
the facts that each run of the quantum computer is almost certain to 
evaluate $r'$ as a divisor of $r$, and that the probability that the 
actual value of $r$ will be found after at most $k$ runs of the 
quantum computer is greater than $1/\zeta(k)$, so that the probability 
is greater than 60\% that no more than 2 runs will be necessary, and 
greater than 90\% that no more than 4 runs will be necessary.
On the other hand, the quantum computer requires more space and time 
to run the refinement (the space and time requirements are each multiplied 
by approximately a constant), and the Quantum Fourier Transform requires 
more delicate rotations of angles (of the order of $\frac{\pi}{n^3}$,
rather than the order of $\frac{\pi}{n^2}$, which is all that Shor's original 
algorithm would require).  Also, the number of runs needed by Shor's original 
algorithm is $O(\log \log n)$, and $\log \log n$ is a very slowly growing 
function.  For a value of $n = 10^{400}$, if the logarithms are to base 2, 
$\log \log n$ is between 10 and 11.  These are questions which will have to be 
investigated in greater detail if the case of which algorithm is preferable 
is to be decided.

\section{Acknowledgements}

I would like to thank Ming Yung and Tim Baker for many helpful suggestions.

I would also like to thank Tony Bracken and Michael Nielsen for their
assistance.

\appendix

\section{Proof of \thmref{modmet}} \applabel{modmet}

\bp There are three conditions to be checked in order to show that $\rho_q$ is
a metric.
\ben
\item Note that $|x-y| \geq 0$ for all $x,y \in I_q$.
Since $0 \leq x \leq q-1$ and $0 \leq y \leq q-1$, then $|x-y| \leq q-1$,
and so $q-|x-y| \geq 1 > 0$, so that for all $x, y \in I_q$,
\be
\rho_q(x,y) = \min(|x-y|,q-|x-y|) \geq 0.
\ee

For all $x \in I_q$, $|x-x| = 0$, so $q-|x-x| = q$, and so $\rho_q(x,x) = 0$.
Conversely, suppose $x,y \in I_q$ and that $\rho_q(x,y) = 0$.
Since $\rho_q(x,y) = \min(|x-y|,q-|x-y|)$, then either $|x-y| = 0$ or
$q-|x-y| = 0$.  If $|x-y| = 0$, then $x = y$.  If $q-|x-y| = 0$, then
$|x-y| = q$, contradicting $|x-y| \leq q-1$.

It follows that $\rho_q(x,y) \geq 0$ for all $x,y \in I_q$, and that
$\rho_q(x,y) = 0$ iff $x = y$.
\item Since $|x-y| = |y-x|$, then $\rho_q(x,y) = \rho_q(y,x)$, so that $\rho_q$
is symmetric.
\item If $\rho_q(x,y) = |x-y|$ and $\rho_q(y,z) = |y-z|$, then
\be
\rho_q(x,z) \leq |x-z| \leq |x-y| + |y-z| = \rho_q(x,y) + \rho_q(y,z).
\ee

If $\rho_q(x,y) = |x-y|$ and $\rho_q(y,z) = q-|y-z|$, then, since
\be
|y-z| \leq |x-y| + |x-z|,
\ee
it follows that
\be
\rho_q(x,z) \leq q-|x-z| \leq q+|x-y|-|y-z| = \rho_q(x,y) + \rho_q(y,z).
\ee
Similarly, if $\rho_q(x,y) = q-|x-y|$ and $\rho_q(y,z) = |y-z|$, then
$\rho_q(x,z) \leq \rho_q(x,y) + \rho_q(y,z)$.

If $\rho_q(x,y) = q-|x-y|$ and $\rho_q(y,z) = q-|y-z|$, then $q-|x-y| \leq |x-y|$
and $q-|y-z| \leq |y-z|$, so that $2 |x-y| \geq q$ and $2 |y-z| \geq q$, and
so $|x-y| \geq \frac{q}{2}$ and $|y-z| \geq \frac{q}{2}$.  There are two
cases.
\bi
\item Case 1 ($0 \leq y < \frac{q}{2}$): Since $|x-y| \geq \frac{q}{2}$,
then $y + \frac{q}{2} \leq x \leq q-1$.  Similarly, $y + \frac{q}{2} \leq z
\leq q-1$.  Since $|x-y| = x-y$ and $|y-z| = z-y$, then $\rho_q(x,y) = q+y-x
\geq q-x > z-x$ and $\rho_q(y,z) = q+y-z \geq q-z > x-z$.  It follows that
\be
\rho_q(x,z) = |x-z| < \rho_q(x,y) + \rho_q(y,z).
\ee
\item Case 2 ($\frac{q}{2} \leq y < q-1$): Since $|x-y| \geq \frac{q}{2}$,
then $0 \leq x \leq y - \frac{q}{2} < \frac{q}{2}$.  Similarly, $0 \leq z \leq
y - \frac{q}{2} < \frac{q}{2}$.  Since $|x-y| = y-x$ and $|y-z| = y-z$, then
$\rho_q(x,y) = q+x-y > x \geq x-z$ and $\rho_q(y,z) = q+z-y > z \geq z-x$.  It
follows that
\be
\rho_q(x,z) = |x-z| < \rho_q(x,y) + \rho_q(y,z).
\ee
\ei

It follows that the Triangle Inequality holds.
\een
It follows from these three facts that $\rho_q$ is a metric on $I_q$, to be
called the modular metric.

Recall that $|x-y| \leq q-1$ for $x,y \in I_q$.
If $|x-y| \leq \frac{q}{2}$, then $\rho_q(x,y) \leq |x-y| \leq \frac{q}{2}$.
On the other hand, if $\frac{q}{2} \leq |x-y| \leq q-1$, then
$\rho_q(x,y) \leq q-|x-y| \leq \frac{q}{2}$.
In either case, $\rho_q(x,y) \leq \frac{q}{2}$.
\ep

\section{Proof of results presented in \secref{firstreg}} \applabel{firstreg}

The first result to prove is the result that if $\frac{q}{r} \in \BZ$, then
\be
P(X = c) = \left\{ \ba{ll} \frac{1}{r}, & \frac{cr}{q} \in \BZ,\\
0, & \mbox{otherwise}. \ea \right.
\ee

In this case, then
\bse \eqlabel{qr_int}
\left\lfloor\frac{q+k}{r}\right\rfloor = \frac{q}{r},
\ese
for $k = 0,\dots,r-1$.
If $\frac{cr}{q} \in \BZ$, then substitution of \eqref{qr_int} into 
\eqref{cr_int} immediately yields 
\be
P(X = c) = \frac{1}{r}.
\ee
On the other hand, in the case that $\frac{cr}{q} \notin \BZ$, then
$\sin(\frac{\pi cr}{q} \frac{r}{q}) = \sin(\pi c) = 0$ (as $c \in \BZ$), so 
that substitution of \eqref{qr_int} into \eqref{cr_non_int} immediately yields 
\be
P(X = c) = 0.
\ee

The next result is that if $1 \leq q \leq r$, then
\be
P(X = c) = \frac{1}{q},
\ee
for all $c$ (thus yielding no useful information).

In this case,
\be
\left\lfloor\frac{q+k}{r}\right\rfloor = \left\{ \ba{ll}
0, & 0 \leq k < r-q,\\1, & r-q \leq k \leq r-1, \ea \right.
\ee
so that 
\be
P(X = c) &=& \frac{1}{q^2} \sum_{k=0}^{r-1} \left|
\sum_{b=0}^{\left\lfloor\frac{q+k}{r}\right\rfloor-1}
\exp\left(\frac{2\pi ibcr}{q}\right) \right|^2\\
&=& \frac{1}{q^2} \sum_{k=r-q}^{r-1} \left| \sum_{b=0}^{0}
\exp\left(\frac{2\pi ibcr}{q}\right) \right|^2\\
&=& \frac{1}{q^2} \sum_{k=r-q}^{r-1} | 1 |^2\\
&=& \frac{1}{q^2} q\\
&=& \frac{1}{q}.
\ee
Alternatively, \eqref{cr_int} and \eqref{cr_non_int} can be used, and they 
yield the same result.

In the case that $q > r$, then for $k = 0,\dots,r-1$, $q \leq q+k < q+r$, so 
that
\be
\frac{q}{r} - 1 < \left\lfloor\frac{q+k}{r}\right\rfloor < \frac{q}{r} + 1.
\ee

\eqref{non_int_int} is now proven as follows.
If $\frac{cr}{q} \in \BZ$, then it follows from \eqref{cr_int} that 
\be
\frac{1}{q^2} \sum_{k=0}^{r-1} \left(\frac{q}{r} - 1\right)^2
< P(X = c) = \frac{1}{q^2} \sum_{k=0}^{r-1}
\left\lfloor\frac{q+k}{r}\right\rfloor^2
< \frac{1}{q^2} \sum_{k=0}^{r-1} \left(\frac{q}{r} + 1\right)^2,
\ee
and so, expanding the squares,
\be
\frac{1}{r} - \frac{2}{q} + \frac{r}{q^2} < P(X = c)
< \frac{1}{r} + \frac{2}{q} + \frac{r}{q^2}.
\ee

On the other hand, if $\frac{cr}{q} \notin \BZ$, then it follows from 
\eqref{cr_non_int} that
\be
P(X = c) &=& \frac{1}{q^2} \sum_{k=0}^{r-1}
\frac{\sin^2\left(\frac{\pi cr}{q}\left\lfloor\frac{q+k}{r}\right\rfloor
\right)}{\sin^2\frac{\pi cr}{q}}\\
&\leq& \frac{1}{q^2} \sum_{k=0}^{r-1}
\frac{1}{\sin^2\frac{\pi cr}{q}}\\
&=& \frac{r}{q^2 \sin^2\frac{\pi cr}{q}},
\ee
thus demonstrating \eqref{upper_bound}.

Suppose that $c = \frac{dq}{r} + \Delta$, where $d \in \BZ$ and $0 <
| \Delta | < \frac{q}{r}$, then (substituting for $c$ in \eqref{cr_non_int}),
\be
P(X = c) &=& \frac{1}{q^2} \sum_{k=0}^{r-1}
\frac{\sin^2\left(\pi d \left\lfloor\frac{q+k}{r}\right\rfloor
+ \frac{\pi \Delta r}{q} \frac{q}{r}
+ \frac{\pi \Delta r}{q}\left( \left\lfloor\frac{q+k}{r}\right\rfloor
- \frac{q}{r} \right)\right)}{\sin^2\left(\pi d
+ \frac{\pi \Delta r}{q}\right)}\\
&=& \frac{1}{q^2} \sum_{k=0}^{r-1} \frac{\sin^2\left(\pi \Delta 
+ \frac{\pi \Delta r}{q}\left( \left\lfloor\frac{q+k}{r}\right\rfloor
- \frac{q}{r} \right)\right)}{\sin^2 \frac{\pi \Delta r}{q}},
\ee
exploiting the periodicity of $\sin$.

If $\frac{dq}{r} \in \BZ$, then $\Delta \in \BZ$, so that
\be
P(X = c) &=& \frac{1}{q^2} \sum_{k=0}^{r-1} \frac{\sin^2\left(\pi \Delta 
+ \frac{\pi \Delta r}{q}\left( \left\lfloor\frac{q+k}{r}\right\rfloor
- \frac{q}{r} \right)\right)}{\sin^2 \frac{\pi \Delta r}{q}}\\
&=& \frac{1}{q^2} \sum_{k=0}^{r-1} \frac{\sin^2\left(
\frac{\pi \Delta r}{q}\left( \left\lfloor\frac{q+k}{r}\right\rfloor
- \frac{q}{r} \right)\right)}{\sin^2 \frac{\pi \Delta r}{q}}.
\ee
In particular, since
\be
\left| \left\lfloor\frac{q+k}{r}\right\rfloor - \frac{q}{r} \right| < 1,
\ee
for all $k = 0,\dots,r-1$, then if $0 < | \Delta | \leq \frac{q}{2r}$, then
\be
\left| \frac{\pi \Delta r}{q}\left( \left\lfloor\frac{q+k}{r}\right\rfloor
- \frac{q}{r} \right) \right| < \left| \frac{\pi \Delta r}{q} \right|
\leq \frac{\pi}{2},
\ee
and so
\be
P(X = c) = \frac{1}{q^2} \sum_{k=0}^{r-1} \frac{\sin^2\left(
\frac{\pi \Delta r}{q}\left( \left\lfloor\frac{q+k}{r}\right\rfloor
- \frac{q}{r} \right)\right)}{\sin^2\frac{\pi \Delta r}{q}}
< \frac{1}{q^2} \sum_{k=0}^{r-1} 1 = \frac{r}{q^2},
\ee
as $\sin^2 y$ is a monotonic increasing function on $y \in [0,\frac{\pi}{2}]$,
thus yielding \eqref{non_int_int_delta}.
This means that the probability that $X$ will differ from $\frac{dq}{r}$, for
some integer $d$ such that $\frac{dq}{r}$ is also an integer, by at most 
$\frac{q}{2r}$ and by more than $0$, is negligible if $q$ is much larger than 
$r$.

If $\frac{dq}{r} \notin \BZ$, then $\Delta \notin \BZ$.  Since 
\be
\frac{q}{r} - 1 < \left\lfloor\frac{q+k}{r}\right\rfloor < \frac{q}{r} + 1,
\ee
for $k = 0,\dots,r-1$, so that
\be
\left| \left\lfloor\frac{q+k}{r}\right\rfloor - \frac{q}{r} \right| < 1,
\ee
and since $|\cos y| \leq 1$ for $y \in \BR$, it follows that
\be
|\sin(\pi \Delta)| - \frac{\pi |\Delta| r}{q} <
\left| \sin\left(\pi \Delta + \frac{\pi \Delta r}{q}\left(
\left\lfloor\frac{q+k}{r}\right\rfloor - \frac{q}{r} \right)\right) \right|
< |\sin(\pi \Delta)| + \frac{\pi |\Delta| r}{q},
\ee
for $k = 0,\dots,r-1$, thus giving bounds on the square root of the numerator 
of the summand in \eqref{cr_non_int}.  Since $\frac{dq}{r} + \Delta$ is an 
integer, then 
\be
|\sin(\pi \Delta)| = \left| \sin\frac{\pi dq}{r} \right|,
\ee
so that 
\be
\left| \sin\frac{\pi dq}{r} \right| - \frac{\pi |\Delta| r}{q} <
\left| \sin\left(\pi \Delta + \frac{\pi \Delta r}{q}\left(
\left\lfloor\frac{q+k}{r}\right\rfloor - \frac{q}{r} \right)\right) \right|
< \left| \sin\frac{\pi dq}{r} \right| + \frac{\pi |\Delta| r}{q},
\ee
for $k = 0,\dots,r-1$.
Upon taking the square (so we now have the numerator of the summand), it 
follows that
\be
&& \sin^2\left(\pi \Delta + \frac{\pi \Delta r}{q}\left(
\left\lfloor\frac{q+k}{r}\right\rfloor - \frac{q}{r} \right)\right)\\
&<& \left(\left|\sin\frac{\pi dq}{r}\right|
+ \frac{\pi |\Delta| r}{q}\right)^2\\
&=& \sin^2\frac{\pi dq}{r}
+ \frac{2\pi |\Delta| r}{q} \left|\sin\frac{\pi dq}{r}\right|
+ \frac{\pi^2 \Delta^2 r^2}{q^2},
\ee
and that if $|\Delta| \leq \frac{q}{\pi r} |\sin\frac{\pi dq}{r}|$, then
\be
&& \sin^2\left(\pi \Delta + \frac{\pi \Delta r}{q}\left(
\left\lfloor\frac{q+k}{r}\right\rfloor - \frac{q}{r} \right)\right)\\
&>& \left(\left|\sin\frac{\pi dq}{r}\right|
- \frac{\pi |\Delta| r}{q}\right)^2\\
&=& \sin^2\frac{\pi dq}{r}
- \frac{2\pi |\Delta| r}{q} \left|\sin\frac{\pi dq}{r}\right|
+ \frac{\pi^2 \Delta^2 r^2}{q^2},
\ee
and so, substituting into \eqref{cr_non_int},
\bne
P(X = c) &<& \frac{r}{q^2 \sin^2\frac{\pi \Delta r}{q}}
\left(\left|\sin\frac{\pi dq}{r}\right|
+ \frac{\pi |\Delta| r}{q}\right)^2 \nonumber\\
&<& \frac{1}{\pi^2 \Delta^2 r
\left(1 - \frac{\pi^2 \Delta^2 r^2}{6 q^2}\right)^2}
\left(\left|\sin\frac{\pi dq}{r}\right| + \frac{\pi |\Delta| r}{q}\right)^2
\eqlabel{non_int_upper}\\
&=& \frac{1}{\pi^2 \Delta^2 r
\left(1 - \frac{\pi^2 \Delta^2 r^2}{6 q^2}\right)^2}
\left(\sin^2\frac{\pi dq}{r}
+ \frac{2 \pi |\Delta| r \left|\sin\frac{\pi dq}{r}\right|}{q}
+ \frac{\pi^2 \Delta^2 r^2}{q^2}\right) \nonumber\\
&=& \frac{1}{\left(1 - \frac{\pi^2 \Delta^2 r^2}{6 q^2}\right)^2}
\left(\frac{\sin^2\frac{\pi dq}{r}}{\pi^2 \Delta^2 r}
+ \frac{2 \left|\sin\frac{\pi dq}{r}\right|}{\pi |\Delta| q}
+ \frac{r}{q^2}\right), \nonumber
\ene
thus yielding \eqref{delta_upper}, if $|\Delta| \leq \frac{q}{2r}$, since
\be
0 < y \left(1 - \frac{1}{6} y^2\right) < \sin y < y,
\ee
for $y \in (0,\frac{\pi}{2}]$.
Similarly, if $|\Delta| \leq \frac{q}{\pi r} |\sin\frac{\pi dq}{r}|$, then
\bne
P(X = c) &>& \frac{r}{q^2 \sin^2\frac{\pi \Delta r}{q}}
\left(\left|\sin\frac{\pi dq}{r}\right|
- \frac{\pi |\Delta| r}{q}\right)^2 \nonumber\\
&>& \frac{1}{\pi^2 \Delta^2 r}
\left(\left|\sin\frac{\pi dq}{r}\right| - \frac{\pi |\Delta| r}{q}\right)^2
\eqlabel{non_int_lower}\\
&=& \frac{1}{\pi^2 \Delta^2 r}
\left(\sin^2\frac{\pi dq}{r}
- \frac{2 \pi |\Delta| r \left|\sin\frac{\pi dq}{r}\right|}{q}
+ \frac{\pi^2 \Delta^2 r^2}{q^2}\right) \nonumber\\
&=& \frac{\sin^2\frac{\pi dq}{r}}{\pi^2 \Delta^2 r}
- \frac{2 \left|\sin\frac{\pi dq}{r}\right|}{\pi |\Delta| q} + \frac{r}{q^2},
\nonumber
\ene
thus yielding \eqref{delta_lower}.

If $\frac{dq}{r} \in \BZ$, then by \eqref{non_int_int} 
and \eqref{non_int_int_delta},
\be
&& \frac{1}{r} - \frac{2}{q} + \frac{r}{q^2}\\
&<& P\left(X = \frac{dq}{r}\right)\\
&\leq& P\left(\rho_q\left(X,\frac{dq}{r}\right) \leq \frac{q}{2r} \right)\\
&=& P\left( X = \frac{dq}{r} \right) + 
P\left(0 < \rho_q\left(X,\frac{dq}{r}\right) \leq \frac{q}{2r} \right)\\
&<& \frac{1}{r} + \frac{2}{q} + \frac{r}{q^2} +
2 \left\lfloor \frac{q}{2r} \right\rfloor \frac{r}{q^2},
\ee
where $\rho_q$ is the modular metric \eqref{mod},
since $P(X = c) < \frac{r}{q^2}$ for all $c$ such that $0 <
\rho_q(c,\frac{dq}{r}) \leq \frac{q}{2r}$ and
\be
\#\left\{ c : 0 < \rho_q(c,\frac{dq}{r}) \leq \frac{q}{2r} \right\}
= 2 \left\lfloor \frac{q}{2r} \right\rfloor.
\ee
It follows that
\be
&& \frac{1}{r} - \frac{2}{q} + \frac{r}{q^2}\\
&<& P\left(\rho_q\left(X,\frac{dq}{r}\right) \leq \frac{q}{2r} \right)\\
&<& \frac{1}{r} + \frac{2}{q} + \frac{r}{q^2}
+ 2 \frac{q}{2r} \frac{r}{q^2}\\
&=& \frac{1}{r} + \frac{2}{q} + \frac{r}{q^2} + \frac{1}{q}\\
&=& \frac{1}{r} + \frac{3}{q} + \frac{r}{q^2},
\ee
this yielding \eqref{int_full}.

The value of the parameter $q$ will now be restricted so that $q > 2r$.

We are interested in the probability that the measured value $c$ of the 
first register will fall inside a specified distance from an integral 
multiple of $\frac{q}{r}$, so we are also interested in the probability 
that it will fall outside the specified distance.  This is the motivation
behind the following calculations.

Since
\be
P(X = c) \leq \frac{r}{q^2 \sin^2\frac{\pi cr}{q}},
\ee
by \eqref{upper_bound}, if $c \in \BZ$, $0 \leq c \leq q$, and 
$\frac{cr}{q} \notin \BZ$, and since $\sin^2 y$ is a monotonic increasing 
function on $y \in [0,\frac{\pi}{2}]$, then the following hold by 
straightforward substitution.
\bi
\item If $d \in \BZ$, $d \in \{ 0,1,\dots,r-1 \}$, $1 < C \leq \frac{q}{2r}$,
and $\frac{dq}{r} + C \in \BZ$, then
\bne
P\left(X = \frac{dq}{r} + C\right) &\leq& 
\frac{r}{q^2 \sin^2\left(\pi d + \frac{\pi Cr}{q}\right)} \nonumber\\
&=& \frac{r}{q^2 \sin^2\frac{\pi Cr}{q}} \eqlabel{integral1}\\
&<& \int_{C-1}^C \frac{r}{q^2 \sin^2\frac{\pi \xi r}{q}} \, d\xi, \nonumber
\ene
the equality following from the fact that $d \in \BZ$;
\item If $d \in \BZ$, $d \in \{ 1,\dots,r \}$, $1 < C \leq \frac{q}{2r}$,
and $\frac{dq}{r} - C \in \BZ$, then
\be
P\left(X = \frac{dq}{r} - C\right) &\leq& 
\frac{r}{q^2 \sin^2\left(\pi d - \frac{\pi Cr}{q}\right)}\\
&=& \frac{r}{q^2 \sin^2\frac{\pi Cr}{q}}\\
&<& \int_{C-1}^C \frac{r}{q^2 \sin^2\frac{\pi \xi r}{q}} \, d\xi,
\ee
the equality following from the fact that $d \in \BZ$.
\ei

If $d \in \BZ$, $d \in \{ 0,1,\dots,r-1 \}$, $1 < A \leq A' \leq
\frac{q}{2r}$, and $\frac{dq}{r} + A, \frac{dq}{r} + A' \in \BZ$, then
by \eqref{integral1},
\bne
&& P\left(\frac{dq}{r} + A \leq X \leq \frac{dq}{r} + A'\right) \nonumber\\
&\leq& \sum_{c = \frac{dq}{r} + A}^{\frac{dq}{r} + A'}
\frac{r}{q^2 \sin^2\frac{\pi cr}{q}} \nonumber\\
&<& \int_{A - 1}^{A'} \frac{r}{q^2} \csc^2\frac{\pi \xi r}{q} \, d\xi
\eqlabel{integral2}\\
&=& - \frac{1}{\pi q} \left[ \cot\frac{\pi \xi r}{q} \right]_{A - 1}^{A'}
\nonumber\\
&=& \frac{1}{\pi q} \left( \cot\frac{\pi (A - 1) r}{q}
- \cot\frac{\pi A' r}{q} \right). \nonumber
\ene
This gives an upper bound on the probability that $X$ will fall between 
$\frac{dq}{r} + A$ and $\frac{dq}{r} + A'$.

Similarly, if $d \in \BZ$, $d \in \{ 1,\dots,r \}$, $1 < B \leq B'
\leq \frac{q}{2r}$, and $\frac{dq}{r} - B, \frac{dq}{r} - B' \in \BZ$, then
\bse \eqlabel{integral3}
P\left(\frac{dq}{r} - B' \leq X \leq \frac{dq}{r} - B\right)
< \frac{1}{\pi q} \left( \cot\frac{\pi (B - 1) r}{q}
- \cot\frac{\pi B' r}{q} \right).
\ese

If $d \in \BZ$, $d \in \{ 0,1,\dots,r-1 \}$, $1 < A \leq \frac{q}{2r}$, 
$1 < B \leq \frac{q}{2r}$, and $\frac{dq}{r} + A, \frac{(d+1)q}{r} - B \in 
\BZ$, the let
\be
C = \left\lfloor \frac{(2d+1)q}{2r} \right\rfloor - \frac{dq}{r},
\ee
so that $C \leq \frac{q}{2r} < C + 1$, and then, by \eqref{integral2} and
\eqref{integral3},
\be
&& P\left(\frac{dq}{r} + A \leq X \leq \frac{(d+1)q}{r} - B\right)\\
&=& P\left(\frac{dq}{r} + A \leq X \leq \frac{dq}{r} + C\right)
+ P\left(\frac{(d+1)q}{r} - \left(\frac{q}{r} - (C + 1)\right) \leq X \leq
\frac{(d+1)q}{r} - B\right)\\
&<& \frac{1}{\pi q} \left( \cot\frac{\pi (A - 1) r}{q}
- \cot\frac{\pi C r}{q} \right)
+ \frac{1}{\pi q} \left( \cot\frac{\pi (B - 1) r}{q}
- \cot\left(\pi - \frac{\pi (C + 1) r}{q}\right) \right)\\
&=& \frac{1}{\pi q} \left( \cot\frac{\pi (A - 1) r}{q}
+ \cot\frac{\pi (B - 1) r}{q} - \cot\frac{\pi C r}{q}
+ \cot\frac{\pi (C + 1) r}{q} \right)\\
&<& \frac{1}{\pi q} \left( \cot\frac{\pi (A - 1) r}{q}
+ \cot\frac{\pi (B - 1) r}{q} \right),
\ee
since $\frac{\pi C r}{q} \leq \frac{\pi}{2} < \frac{\pi (C + 1) r}{q}$, 
so that $\cot\frac{\pi C r}{q}$ is positive, and 
$\cot\frac{\pi (C + 1) r}{q}$ is negative.
Since $\cot y < \frac{1}{y}$ for $0 < y \leq \frac{\pi}{2}$, then it follows 
that
\be
P\left(\frac{dq}{r} + A \leq X \leq \frac{(d+1)q}{r} - B\right)
< \frac{1}{\pi^2 r} \left( \frac{1}{A - 1} + \frac{1}{B - 1} \right).
\ee
Suppose $0 < u \leq \frac{q}{2r} - 1$, then it follows that
\be
P\left(\frac{dq}{r} + u + 1 \leq X \leq \frac{(d+1)q}{r} - u - 1 \right)
< \frac{1}{\pi^2 r} \left( \frac{1}{u} + \frac{1}{u} \right)
= \frac{2}{\pi^2 r u},
\ee
thus demonstrating \eqref{far}.

Adopting the same definitions of $d_c$ and $\Delta_c$ that were used in 
\secref{firstreg}, then it follows that
\be
P(|\Delta_X| \geq u + 1) = \sum_{d=0}^{r-1} 
P\left(\frac{dq}{r} + u + 1 \leq X \leq \frac{(d+1)q}{r} - u - 1 \right)
< \sum_{d=0}^{r-1} \frac{2}{\pi^2 r u} = \frac{2}{\pi^2 u}.
\ee
Therefore it follows that the probability that the measured value of the 
first register falls outside a specified distance from a multiple of 
$\frac{q}{r}$ is bounded above by a quantity which inversely proportional 
to one less than the distance, with no dependence of the upper bound on the 
size of $q$, thus making the possibility unlikely if the specified distance 
is large.

For each value of $d$, we are interested in the values of
\be
P\left(\rho_q\left(X,\frac{dq}{r}\right) < u + 1\right).
\ee

Upper and lower bounds can be easily determined for
\be
P\left(\rho_q\left(X,\frac{dq}{r}\right) < u + 1\right),
\ee
if $\frac{dq}{r} \in \BZ$, specifically, by \eqref{non_int_int} and
\eqref{non_int_int_delta},
\be
&& \frac{1}{r} - \frac{2}{q} + \frac{r}{q^2}\\
&<& P\left(X = \frac{dq}{r}\right)\\
&\leq& P\left(\rho_q\left(X,\frac{dq}{r}\right) < u + 1\right)\\
&=& P\left( X = \frac{dq}{r} \right) + 
P\left(0 < \rho_q\left( X , \frac{dq}{r} \right) < u + 1 \right)\\
&<& \frac{1}{r} + \frac{2}{q} + \frac{r}{q^2} +
2 \lfloor u + 1 \rfloor \frac{r}{q^2}\\
&\leq& \frac{1}{r} + \frac{2}{q} + \frac{r}{q^2} + \frac{2(u+1)r}{q^2}\\
&=& \frac{1}{r} + \frac{2}{q} + \frac{(2u+3)r}{q^2},
\ee
thus leading to \eqref{int}.

If $\frac{dq}{r} \notin \BZ$, then 
\be
P\left(\left| X - \frac{dq}{r}\right| < u + 1\right)
= \sum_{c \in \BZ \atop \left| c - \frac{dq}{r}\right| < u + 1}
P\left( X = c \right),
\ee
so that, by \eqref{non_int_upper},
\be
&& P\left(\left| X - \frac{dq}{r}\right| < u + 1\right)\\
&<& \sum_{c \in \BZ \atop \left| c - \frac{dq}{r}\right| < u + 1}
\frac{1}{\left(1 - \frac{\pi^2 (u+1)^2 r^2}{6 q^2}\right)^2}
\left( \frac{\sin^2\frac{\pi dq}{r}}{\pi^2 \left(c - \frac{dq}{r}\right)^2 r}
+ \frac{2 \left|\sin\frac{\pi dq}{r}\right|}{\pi
\left|c - \frac{dq}{r}\right| q} + \frac{r}{q^2} \right),
\ee
since $|\Delta_c| < u + 1$ for all $c$ in the sum, and by 
\eqref{non_int_lower},
\be
&& P\left(\left| X - \frac{dq}{r}\right| < u + 1\right)\\
&>& \sum_{c \in \BZ, \ \left| c - \frac{dq}{r} \right| < u + 1
\atop \left| c - \frac{dq}{r} \right| < \frac{q}{\pi r}
\left| \sin\frac{\pi dq}{r} \right|}
\left( \frac{\sin^2\frac{\pi dq}{r}}{\pi^2 (c - \frac{dq}{r})^2 r}
- \frac{2 \left|\sin\frac{\pi dq}{r}\right|}{\pi \left| c - \frac{dq}{r}
\right| q} + \frac{r}{q^2} \right).
\ee

It follows that
\be
&& \sum_{c \in \BZ, \ \left| c - \frac{dq}{r} \right| < u + 1
\atop \left| c - \frac{dq}{r} \right| < \frac{q}{\pi r}
\left| \sin\frac{\pi dq}{r} \right|}
\left( \frac{\sin^2\frac{\pi dq}{r}}{\pi^2 (c - \frac{dq}{r})^2 r}
- \frac{2 \left|\sin\frac{\pi dq}{r}\right|}{\pi \left| c - \frac{dq}{r}
\right| q} + \frac{r}{q^2} \right)\\
&<& P\left(\left| X - \frac{dq}{r}\right| < u + 1\right)\\
&<& \frac{1}{\left(1 - \frac{\pi^2 (u+1)^2 r^2}{6 q^2}\right)^2}
\sum_{c \in \BZ \atop \left| c - \frac{dq}{r} \right| < u + 1}
\left( \frac{\sin^2\frac{\pi dq}{r}}{\pi^2 (c - \frac{dq}{r})^2 r}
+ \frac{2 \left|\sin\frac{\pi dq}{r}\right|}{\pi \left| c - \frac{dq}{r}
\right| q} + \frac{r}{q^2} \right).
\ee
Specifically, if
\bse \eqlabel{smaller_u}
u \leq \frac{q}{\pi r} \left| \sin\frac{\pi dq}{r} \right| - 1,
\ese
then
\bne
&& \sum_{c \in \BZ \atop \left| c - \frac{dq}{r} \right| < u + 1}
\left( \frac{\sin^2\frac{\pi dq}{r}}{\pi^2 (c - \frac{dq}{r})^2 r}
- \frac{2 \left|\sin\frac{\pi dq}{r}\right|}{\pi \left| c - \frac{dq}{r}
\right| q} + \frac{r}{q^2} \right)\nonumber\\
&<& P\left(\left| X - \frac{dq}{r}\right| < u + 1\right)\nonumber\\
&<& \frac{1}{\left(1 - \frac{\pi^2 (u+1)^2 r^2}{6 q^2}\right)^2}
\sum_{c \in \BZ \atop \left| c - \frac{dq}{r} \right| < u + 1}
\left( \frac{\sin^2\frac{\pi dq}{r}}{\pi^2 (c - \frac{dq}{r})^2 r}
+ \frac{2 \left|\sin\frac{\pi dq}{r}\right|}{\pi \left| c - \frac{dq}{r}
\right| q} + \frac{r}{q^2} \right). \eqlabel{near}
\ene

The terms inside the sums on the upper and lower bounds in \eqref{near} will 
now be investigated, one at a time.

By the Mittag-Leffler expansion into partial fractions for $\csc^2(\pi z)$ 
from complex analysis (which can be found in many books on complex analysis,
such as \cite{remm,moore}, or by differentiating the Mittag-Leffler expansion 
for $\cot(\pi z)$, which can also be found in books on complex analysis, such 
as \cite{beren,fuchs}),
\be
\sum_{n=-\infty}^\infty \frac{1}{\pi^2 (z-n)^2} = \frac{1}{\sin^2(\pi z)},
\ee
it follows that
\be
\sum_{c \in \BZ} \frac{\sin^2\frac{\pi dq}{r}}{\pi^2 (c - \frac{dq}{r})^2 r}
= \frac{1}{r},
\ee
so that 
\be
\sum_{c \in \BZ \atop \left| c - \frac{dq}{r} \right| < u + 1}
\frac{\sin^2\frac{\pi dq}{r}}{\pi^2 (c - \frac{dq}{r})^2 r}
= \frac{1}{r}
- \sum_{c \in \BZ \atop \left| c - \frac{dq}{r} \right| \geq u + 1}
\frac{\sin^2\frac{\pi dq}{r}}{\pi^2 (c - \frac{dq}{r})^2 r}.
\ee
For $z \in \BR \backslash \BZ$,
\be
\sum_{n \in \BZ \atop n-z \geq u+1} \frac{1}{(n-z)^2} 
&=& \sum_{n \in \BZ \atop n \geq z+u+1} \frac{1}{(n-z)^2}\\
&=& \sum_{n = \lceil z+u+1 \rceil}^\infty \frac{1}{(n-z)^2}\\
&<& \int_{\lceil z+u \rceil}^\infty \frac{1}{(\xi-z)^2} \, d\xi\\
&=& - \left[\frac{1}{\xi-z}\right]_{\lceil z+u \rceil}^\infty\\
&=& \frac{1}{\lceil z+u \rceil - z}\\
&\leq& \frac{1}{u},
\ee
where for real $y$, $\lceil y \rceil$ is the least integer greater than
or equal to $y$.  Similarly, for $z \in \BR \backslash \BZ$, 
\be
\sum_{n \in \BZ \atop n-z \geq u+1} \frac{1}{(n-z)^2} 
&>& \frac{1}{\lceil z+u+1 \rceil - z}\\
&>& \frac{1}{u+2}.
\ee
It follows that
\be
\frac{1}{u+2} < \sum_{n \in \BZ \atop n-z \geq u+1} \frac{1}{(n-z)^2} 
< \frac{1}{u}.
\ee
Similarly,
\be
\frac{1}{u+2} < \sum_{n \in \BZ \atop n-z \leq -(u+1)} \frac{1}{(n-z)^2} 
< \frac{1}{u},
\ee
so that 
\be
\frac{2}{u+2} < \sum_{n \in \BZ \atop |n-z| \geq u+1} \frac{1}{(n-z)^2} 
< \frac{2}{u},
\ee
and so 
\be
\sum_{c \in \BZ \atop \left| c - \frac{dq}{r} \right| < u + 1}
\frac{\sin^2\frac{\pi dq}{r}}{\pi^2 (c - \frac{dq}{r})^2 r}
= \frac{1}{r}
- \sum_{c \in \BZ \atop \left| c - \frac{dq}{r} \right| \geq u + 1}
\frac{\sin^2\frac{\pi dq}{r}}{\pi^2 (c - \frac{dq}{r})^2 r}
\ee
satisfies
\be
\frac{1}{r} - \frac{2 \sin^2\frac{\pi dq}{r}}{\pi^2 r u}
< \sum_{c \in \BZ \atop \left| c - \frac{dq}{r} \right| < u + 1}
\frac{\sin^2\frac{\pi dq}{r}}{\pi^2 (c - \frac{dq}{r})^2 r}
< \frac{1}{r} - \frac{2 \sin^2\frac{\pi dq}{r}}{\pi^2 r (u+2)},
\ee
since $\frac{dq}{r} \notin \BZ$, and so
\bse \eqlabel{dist_squared}
\frac{1}{r} - \frac{2}{\pi^2 r u}
< \sum_{c \in \BZ \atop \left| c - \frac{dq}{r} \right| < u + 1}
\frac{\sin^2\frac{\pi dq}{r}}{\pi^2 (c - \frac{dq}{r})^2 r}
< \frac{1}{r}.
\ese
This accounts for the first term in the sums in the upper and lower bounds 
in \eqref{near}.

Since $\lceil 2u+1 \rceil \leq \#\{ c \in \BZ : |c-\frac{dq}{r}| < u+1 \}
\leq \lceil 2u+2 \rceil$, then
\bse \eqlabel{constant}
\frac{r (2u+1)}{q^2} \leq \sum_{c \in \BZ \atop
\left| c - \frac{dq}{r} \right| < u + 1} \frac{r}{q^2} < \frac{r (2u+3)}{q^2}.
\ese
This accounts for the third term in the sums in the upper and lower bounds 
in \eqref{near}.

All that remains is the second term in the sums.
Since
\be
&& \sum_{c \in \BZ \atop 0 < c - \frac{dq}{r} < u+1}
\frac{1}{c - \frac{dq}{r}}\\
&=& \frac{1}{\left\lceil \frac{dq}{r} \right\rceil - \frac{dq}{r}}
+ \sum_{c \in \BZ \atop 1 < c - \frac{dq}{r} < u+1}
\frac{1}{c - \frac{dq}{r}}\\
&=& \frac{1}{\left\lceil \frac{dq}{r} \right\rceil - \frac{dq}{r}}
+ \sum_{c = \left\lceil \frac{dq}{r} \right\rceil + 1}^{\left\lfloor u
+ \frac{dq}{r} \right\rfloor + 1} \frac{1}{c - \frac{dq}{r}}\\
&<& \frac{1}{\left\lceil \frac{dq}{r} \right\rceil - \frac{dq}{r}}
+ \int_{\left\lceil \frac{dq}{r} \right\rceil}^{\left\lfloor u + \frac{dq}{r}
\right\rfloor + 1} \frac{1}{\xi - \frac{dq}{r}} \, d\xi\\
&=& \frac{1}{\left\lceil \frac{dq}{r} \right\rceil - \frac{dq}{r}}
+ \left[ \ln(\xi - \frac{dq}{r}) \right]_{\left\lceil \frac{dq}{r}
\right\rceil}^{\left\lfloor u + \frac{dq}{r} \right\rfloor + 1}\\
&=& \frac{1}{\left\lceil \frac{dq}{r} \right\rceil - \frac{dq}{r}}
+ \ln\left(\frac{\left\lfloor u + \frac{dq}{r} \right\rfloor + 1 -
\frac{dq}{r}}{\left\lceil \frac{dq}{r} \right\rceil - \frac{dq}{r}}\right)\\
&\leq& \frac{1}{\left\lceil \frac{dq}{r} \right\rceil - \frac{dq}{r}}
+ \ln\left(\frac{u + 1}{\left\lceil \frac{dq}{r} \right\rceil
- \frac{dq}{r}}\right),
\ee
and similarly,
\be
&& \sum_{c \in \BZ \atop 0 > c - \frac{dq}{r} > -(u+1)}
\frac{1}{\frac{dq}{r} - c}\\
&<& \frac{1}{\frac{dq}{r} - \left\lfloor \frac{dq}{r} \right\rfloor}
+ \ln\left(\frac{\frac{dq}{r} - \left\lceil \frac{dq}{r} - u \right\rceil + 1}
{\frac{dq}{r} - \left\lfloor \frac{dq}{r} \right\rfloor}\right)\\
&<& \frac{1}{\frac{dq}{r} - \left\lfloor \frac{dq}{r} \right\rfloor}
+ \ln\left(\frac{u + 1}{\frac{dq}{r} - \left\lfloor \frac{dq}{r}
\right\rfloor}\right),
\ee
then
\be
&& \sum_{c \in \BZ \atop |c - \frac{dq}{r}| < u+1}
\frac{1}{|c - \frac{dq}{r}|}\\
&<& \frac{1}{\left\lceil \frac{dq}{r} \right\rceil - \frac{dq}{r}}
+ \frac{1}{\frac{dq}{r} - \left\lfloor \frac{dq}{r} \right\rfloor}
+ \ln\left(\frac{\left\lfloor u + \frac{dq}{r} \right\rfloor + 1
- \frac{dq}{r}}{\left\lceil \frac{dq}{r} \right\rceil - \frac{dq}{r}}\right)
+ \ln\left(\frac{\frac{dq}{r} - \left\lceil \frac{dq}{r} - u \right\rceil
+ 1}{\frac{dq}{r} - \left\lfloor \frac{dq}{r} \right\rfloor}\right)\\
&\leq& \frac{1}{\left\lceil \frac{dq}{r} \right\rceil - \frac{dq}{r}}
+ \frac{1}{\frac{dq}{r} - \left\lfloor \frac{dq}{r} \right\rfloor}
+ \ln\left(\frac{u + 1}{\left\lceil \frac{dq}{r} \right\rceil
- \frac{dq}{r}}\right) + \ln\left(\frac{u + 1}{\frac{dq}{r}
- \left\lfloor \frac{dq}{r}\right\rfloor}\right).
\ee
Since $\frac{1}{r} \leq \frac{dq}{r} - \lfloor \frac{dq}{r} \rfloor \leq
\frac{r-1}{r}$ (and equivalently, $\frac{1}{r} \leq
\lceil \frac{dq}{r} \rceil - \frac{dq}{r} \leq \frac{r-1}{r}$, noting that
for $y \in \BR \backslash \BZ$, $\lceil y \rceil - \lfloor y \rfloor = 1$),
then
\be
\frac{1}{\left\lceil \frac{dq}{r} \right\rceil - \frac{dq}{r}}
+ \frac{1}{\frac{dq}{r} - \left\lfloor \frac{dq}{r} \right\rfloor}
\leq r + \frac{r}{r-1} = \frac{r^2}{r-1}.
\ee
Similarly,
\be
\left(\frac{dq}{r} - \left\lfloor \frac{dq}{r} \right\rfloor\right)
\left(\left\lceil \frac{dq}{r} \right\rceil - \frac{dq}{r}\right)
\geq \frac{1}{r} \frac{r-1}{r} = \frac{r-1}{r^2},
\ee
so that 
\be
\ln\left[\left(\frac{dq}{r} - \left\lfloor \frac{dq}{r} \right\rfloor\right)
\left(\left\lceil \frac{dq}{r} \right\rceil - \frac{dq}{r}\right)\right]
\geq \ln\frac{r-1}{r^2},
\ee
and so 
\bse \eqlabel{dist}
\sum_{c \in \BZ \atop |c - \frac{dq}{r}| < u+1}
\frac{1}{|c - \frac{dq}{r}|} < \frac{r^2}{r-1} + \ln\frac{r^2(u+1)^2}{r-1}.
\ese

Gathering all the information about individual terms, it follows that if 
\be
u \leq \frac{q}{\pi r} \left| \sin\frac{\pi dq}{r} \right| - 1,
\ee
then
\be
&& \frac{1}{r} - \frac{2}{\pi^2 r u} - \frac{2}{\pi q} \left(\frac{r^2}{r-1}
+ \ln\frac{r^2(u+1)^2}{r-1}\right) + \frac{r (2u+1)}{q^2}\\
&<& P\left(\left| X - \frac{dq}{r} \right| < u+1\right)\\
&<& \frac{1}{\left(1 - \frac{\pi^2 (u+1)^2 r^2}{6 q^2}\right)^2}
\left(\frac{1}{r} + \frac{2}{\pi q} \left(\frac{r^2}{r-1}
+ \ln\frac{r^2(u+1)^2}{r-1}\right) + \frac{r (2u+3)}{q^2}\right),
\ee
as a consequence of \eqref{near}, \eqref{dist_squared}, \eqref{constant}
and \eqref{dist}, thus yielding \eqref{non_int}.

\section{Proof of Inequalities in the Analysis of the Refinement} 
\applabel{algo_analysis}

If $\frac{dq}{r} \in \BZ$, then by \eqref{int} and the fact that 
$q \geq 2 w n^3$,
\bne
\frac{1}{r} - \frac{1}{w n^3}
&<& \frac{1}{r} - \frac{1}{w n^3} + \frac{r}{4 w^2 n^6} \nonumber\\
&\leq& \frac{1}{r} - \frac{2}{q} + \frac{r}{q^2} \nonumber\\
&<& P\left(\rho_q\left(X,\frac{dq}{r}\right) < w n\right) \eqlabel{_ref1}\\
&<& \frac{1}{r} + \frac{2}{q} + \frac{(2wn+1)r}{q^2} \nonumber\\
&\leq& \frac{1}{r} + \frac{1}{w n^3} + \frac{(2wn+1)r}{4 w^2 n^6} \nonumber\\
&<& \frac{1}{r} + \frac{1}{w n^3} + \frac{1}{2 w n^4}, \nonumber
\ene
the final statement following from the fact that $r < n$, so that 
$r \leq n-1$, and so 
\be
(2wn+1)r \leq (2wn+1)(n-1) = 2wn^2 - 2wn + n - 1 = 2wn^2 - (2w - 1)n - 1.
\ee
This demonstrates \eqref{ref1}.

If $\frac{dq}{r} \notin \BZ$, then, by \eqref{non_int} and the fact that 
$q \geq 2 w n^3$,
\bne
&& \frac{1}{r} - \frac{2}{\pi^2 (wn - 1)} - \frac{1}{\pi w n^3}
\left(n + 1 + \ln\frac{w^2 n^4}{n-1}\right) \nonumber\\
&<& \frac{1}{r} - \frac{2}{\pi^2 r (wn - 1)} - \frac{1}{\pi w n^3}
\left(\frac{r^2}{r-1} + \ln\frac{r^2 w^2 n^2}{r-1}\right) \nonumber\\
&<& \frac{1}{r} - \frac{2}{\pi^2 r (wn - 1)}
- \frac{2}{\pi q} \left(\frac{r^2}{r-1}
+ \ln\frac{r^2 w^2 n^2}{r-1}\right) + \frac{r (2wn - 1)}{q^2} \nonumber\\
&<& P\left(\left| X - \frac{dq}{r} \right| < w n\right) \eqlabel{_ref2}\\
&<& \frac{1}{\left(1 - \frac{\pi^2 w^2 n^2 r^2}{6 q^2}\right)^2}
\left(\frac{1}{r} + \frac{2}{\pi q} \left(\frac{r^2}{r-1}
+ \ln\frac{r^2 w^2 n^2}{r-1}\right) + \frac{r (2wn + 1)}{q^2}\right)
\nonumber\\
&\leq& \frac{1}{\left(1 - \frac{\pi^2 w^2 n^2 r^2}{24 w^2 n^6}\right)^2}
\left(\frac{1}{r} + \frac{1}{\pi w n^3} \left(\frac{r^2}{r-1}
+ \ln\frac{r^2 w^2 n^2}{r-1}\right) + \frac{r (2wn + 1)}{4 w^2 n^6}\right)
\nonumber\\
&=& \frac{1}{\left(1 - \frac{\pi^2 r^2}{24 n^4}\right)^2}
\left(\frac{1}{r} + \frac{1}{\pi w n^3} \left(\frac{r^2}{r-1}
+ \ln\frac{r^2 w^2 n^2}{r-1}\right) + \frac{r (2wn + 1)}{4 w^2 n^6}\right)
\nonumber\\
&<& \frac{1}{\left(1 - \frac{\pi^2}{24 n^2}\right)^2}
\left(\frac{1}{r} + \frac{1}{\pi w n^3} \left(n + 1
+ \ln\frac{w^2 n^4}{n-1}\right) + \frac{1}{2 w n^4}\right). \nonumber
\ene
The first inequality above follows from 
\bi
\item the fact that 
\be
(n+1)(r-1) - r^2 = (n+1)(r-1) - (r+1)(r-1) - 1 = (n-r)(r-1) - 1 \geq 0,
\ee
since $n-r \geq 1$ and $r > 1$ (which is required by the fact that 
$\frac{dq}{r} \notin \BZ$), 
\item the fact that 
\be
n^2(r-1) - r^2(n-1) = (n-r)(nr - n - r) = (n-r)[(n-1)(r-1)-1] 
\geq 1 (2-1) = 1,
\ee
since $r \geq 2$ and $n \geq r+1 \geq 3$, and so 
\be
\frac{n^2}{n-1} > \frac{r^2}{r-1}.
\ee
\ei
This demonstrates \eqref{ref2}.

\section{Proof of \thmref{zeta}} \applabel{zeta}

\bl For each $j \in J$, define the random variable $B_{ij}$ with sample space 
$\{ 0,1,2,\dots,a_j \}$ by setting 
\be
R_i = \prod_{j \in J} p_j^{B_{ij}}.
\ee
Specifically, $B_{ij}$ is the power to which $p_j$ is raised in the 
prime factorization of $R_i$.  Then:
\bi
\item The probability distribution for $B_{ij}$ is given by 
\be
P(B_{ij} = b) = \left\{ \ba{ll} \frac{p_j^b - p_j^{b-1}}{p_j^{a_j}}, 
& b > 0,\\
\\
\frac{1}{p_j^{a_j}}, & b = 0. \ea \right.
\ee
\item $B_{ij}$ for $i = 1,2,3,\dots$, and $j \in J$, are independent random 
variables.
\ei
\el
\bp Since $B_{ij} = b$ iff $b$ is the power to which $p_j$ is raised in the 
prime factorization of $R_i$, then 
\be
P(B_{ij} = b) = P(p_j^b | R_i \wedge p_j^{b+1} \nmid R_i),
\ee
and so since $R_i = s/\gcd(Z_i,s)$, it follows that 
\be
P(B_{ij} = b) &=& P(p_j^{a_j-b} | \gcd(Z_i,s) \wedge
p_j^{a_j-b+1} \nmid \gcd(Z_i,s))\\
&=& \left\{ \ba{l} P(p_j^{a_j-b} | Z_i \wedge p_j^{a_j-b+1} \nmid Z_i),
\ b > 0,\\
\\
P(p_j^{a_j} | Z_i), \ b = 0. \ea \right.
\ee
The number of elements of $\{ 0,1,2,\dots,s-1 \}$ which are divisible 
by $p_j^{a_j-b}$ is $s/p_j^{a_j-b}$ for all $b = 0,1,2,\dots,a_j$, so 
that, since $Z_i$ is uniformly distributed,
\be
P(p_j^{a_j-b} | Z_i) = \frac{1}{p_j^{a_j-b}} = \frac{p_j^b}{p_j^{a_j}},
\ee
and so
\be
P(B_{ij} = b) = \left\{ \ba{l} \frac{p_j^b-p_j^{b-1}}{p_j^{a_j}}, \ b > 0,\\
\\
\frac{1}{p_j^{a_j}}, \ b = 0. \ea \right.
\ee
This proves the required formula for $P(B_{ij} = b)$.

For the independence of $B_{ij}$, one can invoke the Chinese 
Remainder Theorem (as Knill did in \cite{knill}, for example).
Alternatively, one can also take the following approach.
For $j_1,\dots,j_l \in J$, and for $b_m = 0,1,\dots,a_{j_m}$ for 
$m = 1,\dots,l$, then 
\be
&& P(B_{ij_1} = b_1, B_{ij_2} = b_2, \dots, B_{ij_l} = b_l)\\
&=& P(p_{j_m}^{b_m} | R_i \wedge p_{j_m}^{b_m+1}  \nmid R_i, \mbox{ for all }
m = 1,\dots,l )\\
&=& P(p_{j_m}^{a_{j_m}-b_m} | \gcd(Z_i,s) \wedge
p_{j_m}^{a_{j_m}-b_m+1} \nmid \gcd(Z_i,s), \mbox{ for all } m = 1,\dots,l).
\ee
For any divisor $t$ of $s$, then the number of elements of 
$\{ 0,1,\dots,s-1 \}$ which are divisible by $t$ is $\frac{s}{t}$, 
so that, for all $i$,
\be
P(t | Z_i) = \frac{1}{t}.
\ee
It follows that for $b_m = 0,1,\dots,a_{j_m}$, $m = 1,\dots,l$, then
\bse \eqlabel{indep}
P\left(\prod_{l=1}^m p_{j_m}^{a_{j_m}-b_m} | Z_i\right) 
= \prod_{m=1}^l \frac{1}{p_{j_m}^{a_{j_m}-b_m}}
= \prod_{m=1}^l P(p_{j_m}^{a_{j_m}-b_m} | Z_i).
\ese
For $m = 1,\dots,l, $ define $f_m : \{ -1,0,1,\dots,a_{j_m} \} \to \BR$ by
\be
f_m(b) = \left\{ \ba{ll} \frac{1}{p_j^{a_{j_m}-b}}, & b \geq 0,\\
0, & b = -1, \ea \right.
\ee
then 
\be
P\left(p_{j_m}^{a_{j_m}-b} | \gcd(Z_i,s)\right) = f_m(b),
\ee
for $b = -1,0,1,\dots,a_{j_m}$, and so, with the aid of \eqref{indep}, 
\be
P\left(\prod_{m=1}^l p_{j_m}^{a_{j_m}-b_m} | \gcd(Z_i,s)\right)
= \prod_{m=1}^l f_m(b_m),
\ee
for $b_m = -1,0,1,\dots,a_{j_m}$, $m = 1,\dots,l$.  It follows that
\be
&& P(B_{ij_1} = b_1, B_{ij_2} = b_2, \dots, B_{ij_l} = b_l)\\
&=& P(p_{j_m}^{a_{j_m}-b_m} | \gcd(Z_i,s) \wedge
p_{j_m}^{a_{j_m}-b_m+1} \nmid \gcd(Z_i,s), \mbox{ for all } m = 1,\dots,l)\\
&=& \prod_{m=1}^l (f_m(b_m) - f_m(b_m-1))\\
&=& \prod_{m=1}^l P(B_{ij_m} = b_m),
\ee
for $b_m = 0,1,\dots,a_{j_m}$, $m = 1,\dots,l$.  For example, one can use a 
proof by induction on $q$ to demonstrate that for $b_m = 0,1,\dots,a_{j_m}$, 
$m = 1,\dots,q$, and for $b_m = -1,0,1,\dots,a_{j_m}$, $m = q+1,\dots,l$,
\be
&& P\left(F_m(b_m,Z_i) \mbox{ for } m = 1,\dots,q, \mbox{ and } G_m(b_m,Z_i) 
\mbox{ for } m = q+1,\dots,l\right)\\
&=& \prod_{m=1}^q P(F_m(b_m,Z_i)) \prod_{m=q+1}^l P(G_m(b_m,Z_i))\\
&=& \prod_{m=1}^q (f_m(b_m) - f_m(b_m-1)) \prod_{m=q+1}^l f_m(b_m),
\ee
where $G_m(b,Z_i)$ denotes the proposition denoting that
$p_{j_m}^{a_{j_m}-b}$ divides $\gcd(Z_i,s)$, and $F_m(b,Z_i)$ denotes the
proposition $G_m(b,Z_i) \wedge \neg G_m(b-1,Z_i)$, so that $F_m(b,Z_i)$ is
equivalent to the proposition that $p_{j_m}^{a_{j_m}-b}$ divides
$\gcd(Z_i,s)$ and $p_{j_m}^{a_{j_m}-b+1}$ does not divide $\gcd(Z_i,s)$.
It follows that $B_{ij_1}, B_{ij_2}, \dots, B_{ij_l}$ are independent random 
variables.  Since the set $\{ j_1,j_2,\dots,j_l \}$ was arbitrary, and 
since $Z_i$ are independent random variables, it follows that 
$B_{ij}$ for $i = 1,2,\dots$, and for $j \in J$, are independent random 
variables.
\ep
The proof of \thmref{zeta} can now be given.
\bp Let $B_{ij}$ be random variables as in the proof of the Lemma, and 
define random variables $C_{kj}$ by 
\be
C_{kj} = \max(B_{1j},\dots,B_{kj}),
\ee
then
\be
S_k = \prod_{j \in J} p_j^{C_{kj}}.
\ee
Since the sample space for $B_{ij}$ is $\{0,1,\dots,a_j\}$ for all $i, j$,
then the sample space for $C_{kj}$ is also $\{0,1,\dots,a_j\}$ for all
$k, j$.  From the result in the Lemma that
\be
P(B_{ij} = a_j) = \frac{p_j^{a_j}-p_j^{a_j-1}}{p_j^{a_j}} = 1 - \frac{1}{p_j},
\ee
for all $i, j$, then
\be
P(B_{ij} < a_j) = \frac{1}{p_j},
\ee
for all $i, j$, and so
\be
P(C_{kj} < a_j) = P(B_{ij} < a_j \mbox{ for } i = 1,\dots,k)
= \prod_{i=1}^k P(B_{ij} < a_j) = \frac{1}{p_j^k},
\ee
as a consequence of the independence of $B_{1j}, B_{2j}, \dots, B_{kj}$ 
(which follows from the independence of $Z_i$ for $i = 1,\dots,k$).
It follows that
\be
P(C_{kj} = a_j) = 1 - \frac{1}{p_j^k},
\ee
and so
\be
P(S_k = s) = P(C_{kj} = a_j \mbox{ for all } j \in J) 
= \prod_{j \in J} P(C_{kj} = a_j)
= \prod_{j \in J} \left(1 - \frac{1}{p_j^k}\right),
\ee
as a consequence of the independence of $C_{kj}$ for $j \in J$ (which follows
from the independence of $B_{ij}$ for $i = 1,2,\dots,k$ and $j \in J$).
Therefore
\be
P(S_k = s) = \prod_{j \in J} \left(1 - \frac{1}{p_j^k}\right)
> \prod_{p \mbox{\small{ prime}}} \left(1 - \frac{1}{p^k}\right)
= \frac{1}{\zeta(k)},
\ee
for $k \geq 2$.
\ep

\section{Proof of \eqref{ck=r}} \applabel{prob}

A similar method to the proof of \thmref{zeta} can be used.  Let $r$ have
prime factorization
\be
r = \prod_{j \in J} p_j^{a_j},
\ee
where $J$ is some index set, $p_j$ are distinct primes, and $a_j \geq 1$ 
for all $j \in J$.
For each  prime $p$, define the random variable $E_{kp}$, with sample space 
$\{ 0,1,2,\dots \}$, by setting 
\be
B_k = \prod_{p \mbox{\small{ prime}}} p^{E_{kp}}.
\ee
Specifically, $E_{kp}$ is the power to which $p$ is raised in the 
prime factorization of $B_k$.  By similar arguments to the ideal case, 
treated in \secref{prob} and \appref{zeta}, then:
\bi
\item For a finite set $J_0$ of primes, and for non-negative integers 
$b_p$ for $p \in J_0$,
\be
P\left(E_{kp} = b_p \mbox{ for all } p \in J_0\right) 
= O\left(\frac{1}{wn}\right),
\ee
if $b_p > 0$ for some $p$ not dividing $r$, or $b_{p_j} > a_j$ for some 
$j \in J$ such that $p_j \in J_0$;
\item For a finite set $J_0$ of primes, and for non-negative integers 
$b_p$ for $p \in J_0$,
\be
P\left(E_{kp} = b_p \mbox{ for all } p \in J_0\right) 
= \prod_{j \in J \atop p_j \in J_0, \ b_{p_j} > 0} \frac{p_j^{b_{p_j}} 
\left(1 - \frac{1}{p_j}\right)}{p_j^{a_j}} \prod_{j \in J \atop p_j 
\in J_0, \ b_{p_j} = 0} \frac{1}{p_j^{a_j}} \left(1+O\left(\frac{r}{wn}\right)
\right),
\ee
if $b_p = 0$ for all $p$ not dividing $r$, and $b_{p_j} \leq a_j$ for all 
$j \in J$ such that $p_j \in J_0$.
\ei
It follows that for any subset $J_0 \subseteq J$, 
\be
P\left(\mbox{$E_{kp_j} < a_j$ for all $j \in J_0$, and $E_{kp} = 0$ for all 
$p \nmid r$} \right) 
= \prod_{j \in J_0} \frac{1}{p_j} \left(1+O\left(\frac{r}{wn}\right)\right).
\ee
For all $k$ and primes $p$, define the random variable $F_{kp}$ by 
\be
F_{kp} = \max(E_{1p},\dots,E_{kp}),
\ee
so that
\be
C_k = \prod_{p \mbox{\small{ prime}}} p^{F_{kp}}.
\ee
It follows that for a finite set $J_0$ of primes, and for non-negative 
integers $b_p$ for $p \in J_0$,
\be
P\left(F_{kp} = b_p \mbox{ for all } p \in J_0\right) 
= O\left(\frac{k}{wn}\right),
\ee
if $b_p > 0$ for some $p$ not dividing $r$, or $b_{p_j} > a_j$ for some 
$j \in J$ such that $p_j \in J_0$.

Since $A_k$ are independent random variables, then $B_k$ are independent 
random variables, so that for any subset $J_0 \subseteq J$, 
\be
&& P\left(\mbox{$F_{kp_j} < a_j$ for all $j \in J_0$, and $F_{kp} = 0$ 
for all $p \nmid r$}\right)\\ 
&=& P\left(\mbox{For all $i = 1,\dots,k$, $E_{ip_j} < a_j$ for all $j \in
J_0$, and $E_{ip} = 0$ for all $p \nmid r$}\right)\\
&=& \prod_{i=1}^k P\left(\mbox{$E_{ip_j} < a_j$ for all $j \in J_0$, and 
$E_{ip} = 0$ for all $p \nmid r$}\right)\\
&=& \prod_{i=1}^k \prod_{j \in J_0} \frac{1}{p_j} 
\left(1+O\left(\frac{r}{wn}\right)\right)\\
&=& \prod_{j \in J_0} \frac{1}{p_j^k} 
\left(1+O\left(\frac{kr}{wn}\right)\right).
\ee
Therefore, similarly to the idealized case,
\be
&& P(C_k = r)\\
&=& P\left(\mbox{$F_{kp_j} = a_j$ for all $j \in J$, and $F_{kp} = 0$ 
for all $p \nmid r$}\right)\\
&=& \sum_{J_0 \subseteq J} (-1)^{\#(J_0)} P\left(\mbox{$F_{kp_j} < a_j$ for 
all $j \in J_0$, and $F_{kp} = 0$ for all $p \nmid r$}\right)\\ 
&=& \sum_{J_0 \subseteq J} (-1)^{\#(J_0)} \prod_{j \in J_0} \frac{1}{p_j^k} 
\left(1+O\left(\frac{kr}{wn}\right)\right)\\
&=& \prod_{j \in J} \left(1 - \frac{1}{p_j^k}\right) 
+ \prod_{j \in J} \left(1 + \frac{1}{p_j^k}\right) O\left(\frac{kr}{wn}\right).
\ee
For $k$ bounded and greater than or equal to $2$, since 
\be
\prod_{j \in J} \left(1 + \frac{1}{p_j^k}\right) 
< \sum_{n=1}^\infty \frac{1}{n^k} = \zeta(k),
\ee
and since $r < n$, then for $k \geq 2$ and $k$ small,
\be
P(C_k = r) &=& \prod_{j \in J} \left(1 - \frac{1}{p_j^k}\right) 
+ O\left(\frac{1}{w}\right)\\
&=& \prod_{j \in J} \left(1 - \frac{1}{p_j^k}\right) 
+ O(n^{-\epsilon}),
\ee
since $w = n^\epsilon$, thus demonstrating \eqref{ck=r}.

\end{document}